\documentclass{article}
\usepackage{arxiv}

\usepackage[utf8]{inputenc}
\usepackage[T1]{fontenc}
\usepackage{hyperref}
\usepackage{url}
\usepackage{booktabs}
\usepackage{amsfonts}
\usepackage{amsmath}
\usepackage{amssymb}
\usepackage{nicefrac}
\usepackage{microtype}
\usepackage{graphicx}
\usepackage[numbers,sort&compress]{natbib}
\usepackage{doi}
\usepackage{multirow}
\usepackage{array}
\usepackage{xcolor}
\usepackage{makecell}
\usepackage{subcaption}
\usepackage{placeins}
\usepackage[title]{appendix}

\title{AIvilization v0: Toward Large-Scale Artificial Social Simulation with a Unified Agent Architecture and Adaptive Agent Profiles}

\newif\ifuniqueAffiliation
\uniqueAffiliationtrue

\usepackage{authblk}

\author[1,2]{Wenkai Fan$^*$}
\author[1,2]{Shurui Zhang$^*$}
\author[1,2]{Xiaolong Wang}
\author[1,2]{Haowei Yang}
\author[1,2]{Tsz Wai Chan}
\author[1,2]{Xingyan Chen}
\author[1,2]{Junquan Bi}
\author[2]{Zirui Zhou}
\author[2]{Jia Liu}
\author[1]{Kani Chen$^\dagger$}

\affil[1]{The Hong Kong University of Science and Technology}
\affil[2]{Bauhinia AI}

\date{}
\renewcommand{\shorttitle}{}

\makeatletter
\long\def\blfootnote#1{\begingroup\renewcommand{\thefootnote}{}\footnote{#1}\endgroup}
\makeatother

\hypersetup{
    pdftitle={AIvilization v0: Toward Large-Scale Artificial Social Simulation with a Unified Agent Architecture and Adaptive Agent Profiles},
    pdfauthor={Wenkai Fan, Shurui Zhang, Xiaolong Wang, Haowei Yang, Tsz Wai Chan, Xingyan Chen, Junquan Bi, Zirui Zhou, Jia Liu, Kani Chen},
    pdfkeywords={Large language model agents, Multi-agent systems, Artificial society, Social simulation},
}

\begin{document}
\maketitle
\blfootnote{$^*$These authors contributed equally.}
\blfootnote{$^\dagger$Corresponding author: \texttt{makchen@ust.hk}.}
\blfootnote{$^\ddagger$Project website: \url{https://aivilization.ai}}
\begin{abstract}
AIvilization v0 is a publicly deployed large-scale artificial society that couples a resource-constrained sandbox with a unified LLM-agent architecture, aiming to sustain long-horizon autonomy while remaining executable under a rapidly changing environment. To mitigate the tension between goal stability and reactive correctness, keeping long-horizon objectives on course while each action remains valid in a fast-changing shared world, we introduce (i) a hierarchical branch-thinking planner that decomposes life goals into parallel objective branches and uses simulation-guided validation plus tiered re-planning to ensure feasibility; (ii) an adaptive agent profile with dual-process memory that separates short-term execution traces from long-term semantic consolidation, enabling persistent yet evolving identity; and (iii) a human-in-the-loop steering interface that injects long-horizon objectives and short commands at appropriate abstraction levels, with effects propagated through memory instead of brittle prompt overrides. The environment integrates physiological survival costs, non-substitutable multi-tier production, an AMM-based price mechanism, and a gated education--occupation system. In a large-scale public deployment with tens of thousands of agents, high-frequency transactions from the platform's mature phase reveal stable markets that reproduce key stylized facts of real economies and structured wealth stratification driven by education and access constraints. At the agent level, portraits evolve coherently over long horizons, and human steering is associated with measurably larger short-horizon profile updates. Controlled ablation experiments complement the deployment evidence, showing that our agent architecture is robust in multi-objective, long-horizon settings.
\end{abstract}


\section{Introduction}\label{sec:introduction}

Large language models (LLMs) increasingly serve as the reasoning core of autonomous agents that pursue goals over extended horizons, and such agents are now deployed in persistent, shared worlds where they plan, act, remember, and interact with other agents and human users \cite{wang2024survey,park2023generative}. In these settings an agent faces two demands at once. It must stay teleologically stable, holding to coherent multi-day objectives, and it must remain reactively correct as prices move, resources run short, and other agents change the shared state between its decisions. Holding both at once is the core difficulty. A far-sighted plan that ignores the present soon becomes infeasible, and purely reactive behavior never accumulates toward a distant goal.

This difficulty sharpens when the world is tightly coupled. In a society where an agent's different lines of activity feed back on one another, advancing one objective draws down the same finite resources and time that the others depend on, so the objectives cannot be pursued in isolation. An agent then has to hold several interacting goals at once, keep a long-horizon direction while adapting to conditions that shift between its decisions, and still ensure that each concrete action satisfies the preconditions the world enforces. A single rigid plan is brittle here, because one blocked step can invalidate a long sequence, and purely reactive behavior is short-sighted, because it cannot protect investments that pay off only over a long horizon. This is the long-standing tension between strategic scope and tactical flexibility \cite{rao1995bdi,huang2022language}, made acute by coupling.

Worlds with this coupled character are increasingly instantiated as artificial societies of LLM agents, a line of work that has advanced quickly. Generative agents that remember, reflect, and plan produce believable individual and collective behavior \cite{park2023generative}, and later systems scale to large populations and real-user-aligned pools \cite{piao2025agentsociety,yang2024oasis,zhang2025socioverse} and drive economic mechanisms from macroeconomic activity to financial markets \cite{li2024econagent,hashimoto2025agent}. These systems establish that LLM agents can populate rich social worlds. Two gaps remain for the problem above. First, many environments exercise a single subsystem, or couple several of them only loosely, which mutes the pressure that forces an agent to trade off interlocking objectives. Second, evaluation often rewards expressiveness or final task success, whereas reviews of generative agent-based modeling identify validation as the central unmet challenge \cite{larooij2026validation,zeng2026too}, and interactive benchmarks show that aggregate scores hide failures that surface only under realistic, time-sensitive execution \cite{liu2024agentbench,froger2026gaia2,cao2026beyond}. Executability under a world that pushes back, and the component-level evidence needed to verify it, has received comparatively little attention as an explicit design target.

We present AIvilization v0, an LLM-agent architecture together with the deployed artificial society in which it runs. The architecture answers the strategy--reactivity tension with a planning system that turns a long-horizon objective into validated action, and a dual-process memory that carries the agent's evolving identity. A Branch-Thinking Planner factorizes the objective into parallel, dependent branches, such as personal development, production, trading, and social engagement, decomposes each into abstract sub-tasks, and reconciles them through contextual prioritization and a global synthesis step, so that a coherent long-horizon strategy survives competing demands. Before any action reaches the platform, a pre-execution Action Simulator rolls the candidate sequence forward against the platform's rules and repairs it in tiers, escalating to full re-planning only when local fixes fail, so the agent recovers from a blocked step without re-deriving its entire plan. The memory separates fast execution traces from slow semantic integration and consolidates experience into an adaptive profile, so that identity persists and evolves through social interaction, and human steering enters through the same channel as durable agent state. The society these agents inhabit is tightly coupled. Physiological upkeep, non-substitutable multi-tier production, an automated market maker (AMM) whose prices move with every trade, and education-gated occupations form a closed loop in which each decision reshapes the constraints on later ones. Through this coupling, the environment both demands such an architecture and produces aggregate behavior that merits study in its own right.

This paper makes three contributions. The first is an agent architecture for long-horizon autonomy in coupled simulation worlds, unifying hierarchical branch planning, pre-execution simulation with tiered repair, dual-process memory with an evolving profile, and memory-mediated human steering in a single decision loop. The second is a deployed, persistent artificial society whose tightly coupled economy, physiology, production chains, AMM exchange, and education-gated labor serves as a constrained multi-agent testbed and gives rise to emergent socioeconomic structure. The third is an evaluation of both, which separates controlled ablation evidence from deployment-scale observation. At scale the society sustains bounded, non-degenerate activity, spanning 555{,}402 market records across 24 series and 28{,}107 character-state records, and reproduces canonical stylized facts of real markets and stratification, including heavy-tailed commodity returns (median excess kurtosis 137.0), volatility clustering (absolute-return autocorrelation 0.186), and an education--wealth gradient (Spearman $\rho = 0.648$) \cite{cont2001empirical}. Controlled ablations with 80 agents per variant show that branch and objective decomposition are decisive for complex, multi-objective tasks (education gain $+14.51$, Cliff's $\delta = 0.77$), and a trace audit over 10{,}149 planner--simulator episodes ties the simulator to intent-preserving repair. Agents sustain coherent portraits across 484{,}823 recorded transitions and show measurably larger short-horizon profile updates under denser human steering.

The paper is organized as follows. Section~\ref{sec:related-work} positions the work against research on agent architectures and on artificial-society testbeds and evaluation. Section~\ref{sec:problem-setting} formalizes long-horizon autonomy in a coupled multi-agent society. Section~\ref{sec:architecture} details the agent state and the planning, simulation, memory, and steering components, and Section~\ref{sec:environment} specifies the coupled society that drives them. Section~\ref{sec:evaluation-protocol} presents each experiment with its results. Section~\ref{sec:conclusion} concludes with a synthesis of the contributions, the scope of the evidence, and the outlook.

\section{Related Work}\label{sec:related-work}

\subsection{Agent Architectures for Long-Horizon Autonomy}\label{subsec:rw-agent-architectures}

An LLM agent's competence depends heavily on the scaffolding built around the model \cite{wang2024survey,chowa2026language}. Interleaving reasoning with acting, searching over reasoning paths, reflecting on past failures, and calling external tools all extend what a fixed model can do \cite{yao2022react,yao2023tree,shinn2023reflexion,schick2023toolformer}, and much of an agent's behavior can be located in external memory, skill libraries, and execution harnesses \cite{wang2023voyager,xu2026mem}. Because agents that recycle their own condensed experience can drift or degrade, recent work treats memory as an explicit, managed component \cite{shao2025your}, and for long-horizon goals a recurring design prompts the model to decompose abstract objectives top-down into executable sub-tasks \cite{huang2022language}. Beyond competence, an agent that persists in a social world also needs a stable yet revisable identity, which recent accounts frame through the persona a language model enacts and updates over an interaction \cite{shanahan2023role}.

Making agent structure explicit has long been central to multi-agent systems. The belief--desire--intention tradition represents beliefs, goals, and intentions as first-class objects, and later work combines such structured agents with learning and with multi-level explainability \cite{rao1995bdi,stone2000multiagent,yan2025multi}. When several agents share a world, coordination becomes a structural concern in its own right, studied through protocols with provable properties and through measures of whether a group acts as an integrated collective \cite{bollig2026provable,riedl2025emergent}. These threads each address part of one problem, sustaining a coherent and controllable agent whose local choices stay sound over a long horizon.

A related line of work studies how people direct such agents. Human--agent interaction research specifies when and how a system should involve people \cite{nushi2019guidelines} and treats human input as a rich signal, from hybrid human--agent teams and preference-based value learning to long-horizon user modeling \cite{cohen2025framework,dell2026human,liu2026improving,holgado2026learning,duan2026lifesim,chen2026knowu}, with the common finding that effective input shapes an agent's internal state and persists there. AIvilization draws these threads together in a hierarchical architecture that reconciles strategic coherence with local adaptation under the interlocking constraints of a coupled society, and that routes human steering through the same memory and profile so that direction and identity evolve together.

\subsection{Artificial Societies, Emergent Economies, and Agent Evaluation}\label{subsec:rw-llm-social-agents}

Artificial societies of LLM agents form a large and active genre. Generative agents that store, reflect on, and plan from experience produce believable individual and collective behavior \cite{park2023generative}; later systems scale to large populations, social-media dynamics, and real-user-aligned pools \cite{piao2025agentsociety,yang2024oasis,zhang2025socioverse}, and extend toward desire-driven daily activity, mean-field population dynamics, and social-exchange accounts of interaction \cite{wang2025simulating,mi2026mf,wang2025investigating}. A parallel line lets LLM agents drive economic mechanisms, from macroeconomic activity to financial-market and disclosure dynamics \cite{li2024econagent,hashimoto2025agent,zhao2026green}. These efforts connect to the longer tradition of agent-based computational economics, which grows aggregate regularities from local interaction \cite{tesfatsion2002agent,farmer2009economy} and treats the reproduction of empirical stylized facts, such as heavy tails and volatility clustering, as a test of a synthetic market's realism \cite{cont2001empirical}.

Turning such environments into evidence requires principled testbeds and evaluation. Reproducible testbeds formalize planning and learning under partial observability, cooperative path finding, and adaptation to changing partners and tasks \cite{schwartz2025posggym,phan2026confidence,rother2025open}, and discrete-event methods support the systematic testing of multi-agent systems \cite{baiardi2026testing}. Interactive and asynchronous benchmarks show that even strong models complete only a fraction of multi-step tasks \cite{liu2024agentbench,zhou2024webarena,froger2026gaia2}. Procedure and generality-aware evaluation shows that final task success can conceal time-sensitive or procedurally invalid behavior \cite{cao2026beyond,li2026benchmark}, which motivates tying evaluation to individual architectural components \cite{souza2026toward}. Reviews of generative agent-based modeling similarly identify validation and calibration as the central challenge \cite{larooij2026validation,zeng2026too,fachada2026can}. AIvilization contributes a tightly coupled artificial society whose emergent market and stratification behavior is checked against these stylized facts, together with a component-level evaluation of the architecture that operates within it.

\section{Problem Setting}\label{sec:problem-setting}

AIvilization is a deployed artificial society in which LLM-controlled agents pursue long-horizon goals inside a resource-constrained platform. Its subsystems are tightly coupled, so an agent's choices carry real costs and reshape the conditions that its later choices must satisfy. The problem this paper addresses is turning a natural-language goal into behavior that stays valid under current physiological, economic, institutional, and social conditions, while the agent's own actions draw down shared resources and other agents alter the shared state between its decisions.
We study this as long-horizon autonomy in a multi-agent world. The platform serves as an architecture-oriented testbed.

Let $\mathcal{M}=(\mathcal{A},\mathcal{X},\mathcal{U},F,\mathcal{R})$ denote the environment, where $\mathcal{A}=\{a_1,\ldots,a_n\}$ is the set of agents, $\mathcal{X}$ the global state space, $\mathcal{U}$ the platform action space, $F$ the platform transition rule, and $\mathcal{R}$ the explicit platform rules. Time is continuous game-time $t\ge 0$, and the state at any instant is
\[
x(t)=\big(s_1(t),\ldots,s_n(t),z(t)\big),
\]
where $s_a(t)$ is the state of agent $a$ and $z(t)$ collects the shared platform variables, including market pools, commodity prices, production recipes, and occupation thresholds.

Agents act asynchronously. Each agent $a$ decides at its own increasing sequence of decision times $t^{a}_{0}<t^{a}_{1}<\cdots$, set by its own execution, including the duration of the action it commits, and events such as completion, failure, or a replanning trigger. At a decision time $t^{a}_{k}$, agent $a$ observes $\big(s_a(t^{a}_{k}),z(t^{a}_{k})\big)$ and commits an action $u\in\mathcal{U}_a$; the platform applies this single event through the rule $x\big((t^{a}_{k})^{+}\big)=F\big(x(t^{a}_{k}),a,u\big)$, which updates $s_a$ and the shared $z$ and leaves the other agents' states unchanged until they next act. The event log interleaves all agents' events in game-time order.

The society is constrained by operational mechanisms that together determine which actions an agent can execute at any moment. Physiological state is consumed and restored by action and bounds sustained activity. Production follows fixed recipes with non-substitutable inputs, so an output requires every prerequisite material together with sufficient energy, satiety, and labor time. Trade is mediated by automated market makers whose prices move with each finite transaction, so one agent's trades change the opportunities available to others. Education and occupation access are gated by accumulated study, residential tier, and population-level competition. Each system imposes a precondition that a candidate action must satisfy to be executable. Detail mechanisms, parameters and state transitions will be introduced in  Section~\ref{sec:environment}.

These mechanisms together define the feasible action set
\begin{align*}
\mathcal{U}^{\mathrm{feas}}_a(x(t))
= \{u\in\mathcal{U}_a:\;&
C_{\mathrm{phys}}(u,x(t))=1,\;
C_{\mathrm{inv}}(u,x(t))=1, \nonumber\\
&
C_{\mathrm{market}}(u,x(t))=1,\;
C_{\mathrm{access}}(u,x(t))=1\}.
\end{align*}
Each $C\colon\mathcal{U}\times\mathcal{X}\to\{0,1\}$ is a binary feasibility predicate. $C(u,x(t))=1$ when action $u$ satisfies that constraint in state $x(t)$ and $0$ otherwise, so $\mathcal{U}^{\mathrm{feas}}_a(x(t))$ is the set of actions that pass all four checks at once. The predicates are evaluated at the state $x(t^{a}_{k})$ that agent $a$ observes at its decision. Because other agents act in between, the shared state agent $a$ faces at its next decision generally differs from the one it planned against. Prices, supplies, and institutional conditions may all have moved, so an action that was feasible when planned can be infeasible when executed. Language-level planning therefore should be coupled to mechanisms that preserve executability as the shared state moves.

Within this setting, an agent pursues its current long-horizon objective, which may originate from platform initialization, autonomous reflection, or human steering and is stored in long-term memory instead of a transient prompt, by choosing actions that advance the objective while staying inside the feasible set as the shared state evolves.

\section{Agent Architecture}\label{sec:architecture}

Section~\ref{sec:problem-setting} framed the agent's task as holding a long-horizon strategy together while every action stays feasible in a world that other agents keep changing. Planners typically polarize toward one side of this trade-off, producing either rigid far-sighted plans that break under environmental shifts, or myopic reactive behavior that never reaches an overarching goal. AIvilization addresses the trade-off with a unified architecture built on three mutually reinforcing mechanisms. Hierarchical branch-thinking provides parallel, temporally abstract goal decomposition and planning. Social-cognitive evolution updates the agent's identity through interaction-driven feedback. Dual-process memory consolidation separates fast execution traces from slow, identity-level integration. Together these mechanisms let an agent hold a stable strategic direction while adapting continuously to shifting market signals, physiological constraints, and social dynamics. The
rest of this section details each part.

\subsection{Agent State}\label{subsec:reasoning-cycle}

Each agent is represented by a state
\[
s_a(t)=\big(\mathrm{dyn}_a(t),\mathrm{task}_a(t),\mathrm{mem}_a(t),\mathrm{profile}_a(t)\big),
\]
which holistically describes the agent at any moment. The dynamic state $\mathrm{dyn}_a(t)$ holds the transient attributes that drive immediate decisions, including energy, satiety, health, education score, balance $B_a(t)$, residential tier $R_a(t)$, occupation, and inventory $Q_a(t)$. The task state $\mathrm{task}_a(t)$ records the current long-horizon objective $g_a$, the active branches, the selected sub-task, and the candidate action sequence. The memory state divides into a short-term and a long-term store, $\mathrm{mem}_a(t)=(\mathrm{STM}_a(t),\mathrm{LTM}_a(t))$, which serve immediate execution and enduring identity respectively. The profile $\mathrm{profile}_a(t)$ is the agent's persistent but revisable portrait, initialized from lightweight user-provided role information, including an MBTI-style category used as an interpretable seed, and expanded into the textual fields described in Section~\ref{subsec:memory-profile}. 

\subsection{Hierarchical Planning Structure}\label{subsec:planning-structure}

Agent planning must reconcile long-term strategic coherence with short-term adaptive agility. The planning structure is designed to operate within a complex, rule-governed world and to manage a wide scope of tasks, from individual development such as health and education to multi-agent social dynamics. It achieves this through three components that work as one pipeline, a Branch-Thinking Planner for hierarchical goal management, an Action Simulator for pre-execution validation, and an adaptive re-planning module for robust failure recovery.

\begin{figure}[t]
\centering
\includegraphics[width=\textwidth]{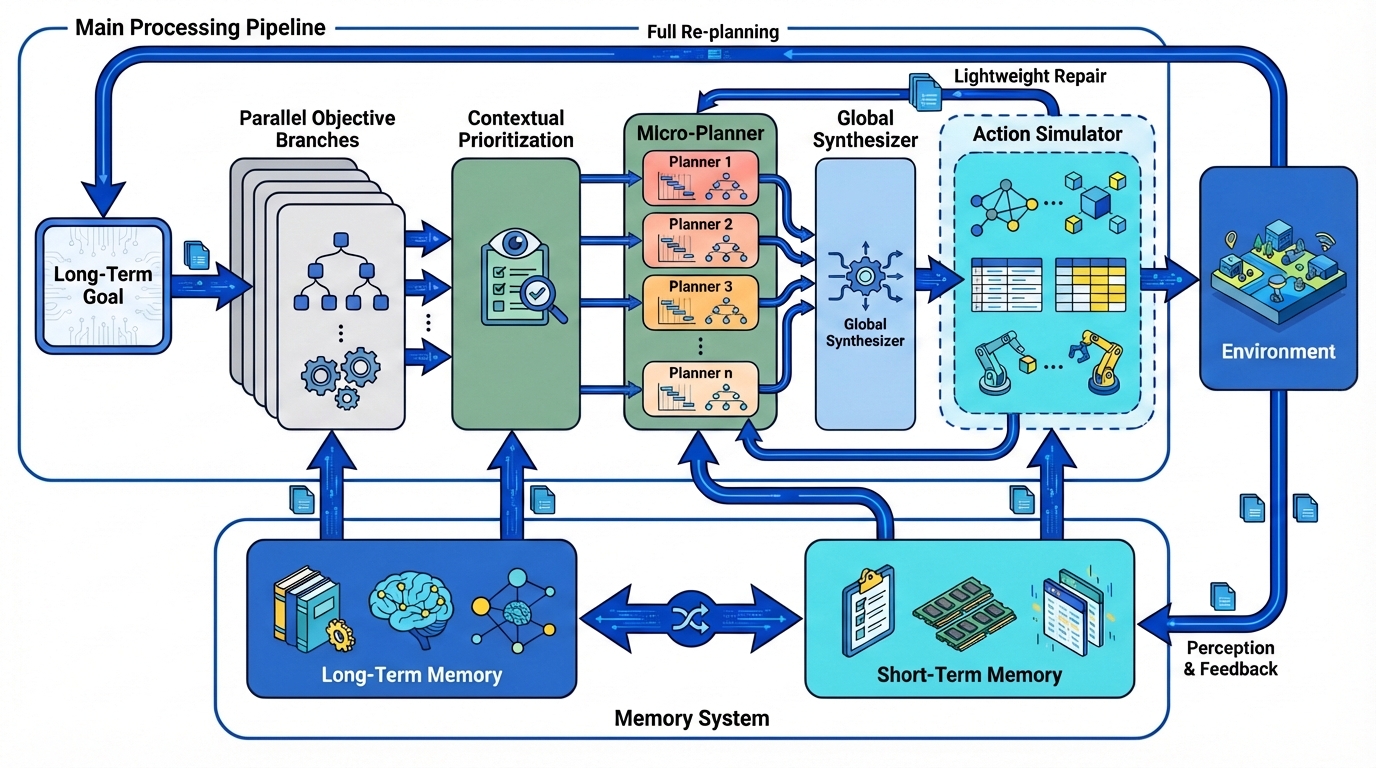}
\caption{Agent cognitive architecture with hierarchical planning, simulation-based filtering, and dual-process memory. Long-term and short-term memory jointly provide guidance and working context to a hierarchical planner, which decomposes a long-term goal into parallel objective branches, prioritizes their sub-tasks by current context, expands the selected sub-task through a domain-specific micro-planner, and reconciles competing branches in a global synthesizer. Proposed actions are evaluated by an Action Simulator that detects infeasible or conflicting behavior, enabling lightweight repair or full re-planning before execution in the environment. Environmental feedback and execution outcomes update short-term state and consolidate long-term experience.}
\label{fig:planning-system}
\end{figure}

\subsubsection{Branch-Thinking Planner}\label{subsec:btp}

At the heart of the framework is the Branch-Thinking Planner, which decomposes the planning horizon into a multi-layered hierarchy spanning objective, branch, sub-task, and action, shown along the top of Figure~\ref{fig:planning-system}. This design tackles the trade-off between strategic vision and tactical flexibility directly.

\paragraph{Strategic decomposition.} At the highest level the agent holds a long-term objective $g_a$, which is strategically decomposed into multiple parallel, semi-independent objective branches, such as personal development, production and resource management, trading and market analysis, and social engagement. By distributing the overarching goal across parallel branches, the planning problem is effectively factorized. Each branch then operates within a focused, constrained subspace, which significantly mitigates the error propagation common in long sequential plans and allows concurrent reasoning about disparate goals, a necessity for managing the multifaceted life of an agent.

Within each individual branch, a second layer of decomposition breaks the branch objective into a sequence of abstract sub-tasks. This hierarchical structure introduces temporal abstraction and grants tactical flexibility, since the agent can adapt its immediate sub-task to real-time feedback and local state constraints without re-evaluating the entire strategic plan. The highest-level decomposition is not invoked at every planning cycle. It is triggered only when a significant shift in environmental conditions or a strategic reassessment warrants it, which preserves computational resources for the real-time adaptation required at lower levels.

\paragraph{Contextual prioritization and selection.} Each planning cycle commences with contextual prioritization and selection. The agent evaluates all active sub-tasks from its parallel objective branches against its real-time state and the current environmental context. The state includes internal attributes drawn from $\mathrm{dyn}_a(t)$, such as health, energy, satiety, balance, and inventory, while the context includes external factors carried in the shared state $z(t)$, such as market prices and world rules. Given the dynamic nature of the world, this step ranks candidates continuously as conditions change. The most viable sub-task, ranked against the agent's long-term objective $g_a$ together with the personality and values held in $\mathrm{profile}_a(t)$, is selected for execution in the current cycle. This grounds the long-term strategy in present reality and prevents wasted effort on obsolete or infeasible actions.

\paragraph{Action sequence generation.} Following selection, the action-generation layer translates the abstract sub-task into a concrete sequence of atomic actions $U_a=(u_{a,1},\ldots,u_{a,m})$. The translation is performed by a domain-specific micro-planner operating within the highly constrained context of the selected branch and target sub-task. Because the preceding layers have already handled strategic decomposition and contextual selection, the micro-planner's potential action space is significantly reduced, and the resulting sequence is more executable and contextually coherent, aligning the agent's immediate operations with both its current state and its broader strategic intent. This structured top-down refinement from abstract goals to concrete actions is fundamental to the framework's ability to be strategic and reactive at once.

\paragraph{Global synthesis.} Finally, the action sequences from the active branches are integrated by the global synthesizer. This is an orchestration process that resolves inter-branch conflicts and ensures global coherence. The module strategically interleaves actions by assigning priorities determined by their contribution to the long-term objective, the urgency of the originating branch, and adherence to shared resource constraints such as time, money, materials, energy, and satiety. This step explicitly addresses the tension between local and global optima. It prevents an over-focus on a single branch's objective from jeopardizing the agent's overall viability or its progress toward the main goal. By re-aligning the integrated plan with the strategic intent established at the highest level, the planner forms a closed loop, so that every executed action remains strategically sound while being tactically chosen.

\subsubsection{Pre-execution Simulation}\label{subsec:action-simulator}

To bridge abstract planning and reliable execution in a dynamic world, the Action Simulator serves as a pre-execution validation layer within the planning structure, designed to enhance plan robustness and mitigate failures arising from unforeseen state changes. Before committing any action sequence to the environment, the simulator performs a counterfactual simulation, projecting the sequence's outcomes by prospectively evolving the agent's state under a predictive model of the world. For a candidate sequence $U_a=(u_{a,1},\ldots,u_{a,m})$ it starts from $\tilde{s}^{(0)}_a=s_a(t)$ and applies the provisional transitions
\[
\tilde{s}^{(\ell)}_a
=\operatorname{Sim}\big(\tilde{s}^{(\ell-1)}_a,u_{a,\ell},z(t),\mathcal{R}\big),
\qquad \ell=1,\ldots,m .
\]
At each provisional step it applies the same feasibility checks the platform enforces, covering physiological viability, material sufficiency, budget and market availability, access or occupation eligibility, and action-specific logic. This forward simulation identifies potential execution failures such as resource shortfalls, constraint violations, and logical inconsistencies in the plan, and it returns the first violated step so that repair can target the specific source of infeasibility.

Upon detecting a potential failure, a two-stage adaptive repair process is initiated. A local repair mechanism first attempts to resolve the issue using a set of predefined, computationally inexpensive heuristics, such as inserting a sleep or eat action, reducing a trade quantity, or acquiring a missing prerequisite. If these simple fixes prove insufficient, the system escalates to a reactive correction module, which leverages knowledge cached in $\mathrm{STM}_a(t)$ from past successes and failures and performs rapid, pattern-based reasoning to generate a viable alternative action or a minor plan modification. The entire simulation and repair loop operates efficiently, bypassing the need to re-invoke the full multi-layered planning stack. This design keeps plan executability high while preserving computational resources, enabling the agent to be deliberate in planning and agile in execution. Table~\ref{tab:as-failure-taxonomy} summarizes the failure signals the simulator detects, the first response each triggers, and the condition under which the system escalates to full re-planning.

\begin{table}[t]
\centering
\caption{Failure signals handled by the Action Simulator and adaptive re-planning.}
\label{tab:as-failure-taxonomy}
\footnotesize
\begingroup
\setlength{\tabcolsep}{4pt}
\renewcommand{\arraystretch}{1.15}
\hyphenpenalty=10000
\begin{tabular}{@{}>{\raggedright\arraybackslash}p{0.15\textwidth}>{\raggedright\arraybackslash}p{0.29\textwidth}>{\raggedright\arraybackslash}p{0.24\textwidth}>{\raggedright\arraybackslash}p{0.24\textwidth}@{}}
\toprule
\textbf{Signal} & \textbf{Typical source} & \textbf{First response} & \textbf{Escalation condition} \\
\midrule
Resource shortfall & Missing materials, insufficient balance, exhausted commodity pool & Heuristic repair or alternate acquisition step & Repeated failure to obtain prerequisite \\
\midrule
Physiological violation & Low energy, satiety, or health before work, study, or production & Insert eat, sleep, or health recovery action & Recovery blocks the branch goal or repeats \\
\midrule
Eligibility violation & Education, occupation, residential tier, or location precondition not met & Choose a reachable sub-task or a study step & Goal requires an inaccessible tier \\
\midrule
Logical inconsistency & Action order conflicts with recipe or platform state & Reorder or replace the local sequence & Same branch remains inconsistent \\
\midrule
Environment shift & Price, inventory, social state, or rule-relevant change before execution & Reactive correction from STM and cached experience & Branch priority or strategy becomes invalid \\
\bottomrule
\end{tabular}
\endgroup
\end{table}

\subsubsection{Adaptive Re-planning and Feedback Loops}\label{subsec:replanning}

Recognizing that no plan survives contact with reality, the framework is equipped with a tiered adaptive re-planning module that handles execution failures and unexpected opportunities while economizing computation. Upon a minor action failure, the agent first attempts a rapid fix. It consults $\mathrm{STM}_a(t)$, a buffer of recent succeeded and failed actions, outcomes, and reflections, to re-evaluate and adjust its immediate next steps. This approach, akin to muscle memory, provides swift and low-cost adaptation for common setbacks. Only when lightweight repairs fail consecutively, or a major environmental shift occurs, does the agent escalate to a full re-plan, invoking the entire Branch-Thinking pipeline to reassess objectives and generate a new strategy from the top down. While costly, this serves as a guaranteed recovery mechanism. The tiered design embodies the framework's balance of responsiveness and strategic deliberation, defaulting to fast fixes and reserving costly re-planning for significant disruptions.

\subsection{Adaptive Agent Profile}\label{subsec:adaptive-profile}

An agent's identity is an adaptive profile that evolves through lived experience, particularly social interaction. This profile serves as the central nexus, integrating the agent's physical state, cognitive processes, and social identity, and its evolution is driven by a continuous cycle of action, social engagement, and memory consolidation. We detail the primary engine of change, social interaction, followed by the cognitive machinery that processes it.

\subsubsection{Social Interaction as the Engine of Evolution}\label{subsec:social-evolution}

Agents in the simulation are social beings whose growth is fundamentally driven by a structured cycle of social engagement. The process begins with proactive planning, where an agent selects interaction targets and conversation topics based on its current goals, state, economic needs, and existing profile. The interaction itself unfolds through natural-language dialogue, facilitating the exchange of knowledge, opinions, and social cues.

Each social event triggers a post-interaction reflection phase. During this phase the agent updates its internal models of other agents, such as relation and attitude, and feeds the experiential data into its memory systems. This act of consolidation is what translates transient social encounters into lasting changes in the agent's core identity, such as its values and behavioral tendencies. This feedback loop, in which the current profile guides social action and social outcomes reshape the profile, is the core mechanism of agent evolution.

\subsubsection{Dual-Process Memory Architecture}\label{subsec:memory-profile}

To manage the distinct demands of immediate action and long-term growth, memory is implemented as a dual-process architecture. This design functionally separates transient, action-oriented experience from enduring, identity-shaping knowledge, mirroring the cognitive distinction between fast, intuitive reasoning and slow, deliberative integration.

\paragraph{Short-term Memory.} Short-term memory $\mathrm{STM}_a(t)$ operates as a high-frequency buffer focused on immediate task execution and outcome evaluation. It continuously records traces from the planner and execution feedback from the environment, cataloguing them as successful actions and failed actions together with simulator decisions and local reflections. When a planned action sequence concludes, whether through full success or by triggering a re-planning cycle, the execution traces are summarized into concrete experiences that serve a direct, pragmatic function. Successful patterns refine the mapping from sub-goals to executable actions, enhancing the efficiency of future planning cycles, while failure patterns prime the re-planning module for quicker recovery from similar setbacks. STM thus embodies a form of fast thinking, allowing the agent to adapt its behavior dynamically to recent outcomes without altering its core identity.

\paragraph{Long-term Memory.} Long-term memory $\mathrm{LTM}_a(t)$ functions as the agent's stable yet evolving semantic core. It integrates significant experiences, particularly those from social interaction, to form a coherent agent profile, which holds structured fields for belief, values, personality, mood, habits, and diary; Appendix~\ref{app:agent-profile} shows an example profile from the deployed platform. Unlike the event-specific logs of STM, LTM performs a slow, integrative synthesis, and the resulting profile acts as a high-level control signal that biases strategic decisions such as goal selection in the contextual prioritization layer. Reflection routines receive the current portrait, recent STM traces, social summaries, and diary context, and revise the portrait after meaningful experience, with the diary linking event-level memory to identity-level revision.

These two memory streams engage in a continuous, bidirectional dialogue. Tactical successes logged repeatedly in STM can gradually consolidate into LTM as new knowledge or stable preferences. The high-level patterns stored in LTM, such as values and personality, provide a top-down cognitive context that shapes how experiences in STM are interpreted and prioritized. This synergy creates a powerful feedback loop in which immediate actions inform long-term identity, and identity in turn guides future actions, so the portrait is a working part of the architecture that conditions planning and is itself revised by experience.

\subsection{Human-in-the-Loop Steering}\label{subsec:human-steering-architecture}

A distinguishing feature of AIvilization is that agents are not deployed as sealed autonomous entities. The platform is explicitly designed for hybrid autonomy, in which humans steer agents in real time while agent-level coherence and executability are preserved. Three complementary intervention channels are deliberately mapped to different levels of the architecture and propagated through different pathways.

\paragraph{Strategic steering.} Users may assign or revise an agent's long-horizon objective $g_a$. In the architecture this input is treated as a top-level goal and integrated into $\mathrm{LTM}_a(t)$. It reshapes the agent's global planning landscape by affecting how the Branch-Thinking Planner constructs and weights parallel branches, how the contextual prioritization layer ranks candidate sub-tasks under resource constraints, and how global synthesis resolves trade-offs across domains. Long-horizon objectives are therefore not implemented as isolated prompts; they are integrated into the full hierarchical planning stack and influence both multi-turn consistency and per-cycle task selection.

This design provides two main advantages. First, it enables stable and interpretable governance, since human intent is expressed at the same abstraction level as the agent's strategic decomposition, which reduces the brittleness that arises when users attempt to steer behavior through repeated low-level instructions. Second, it yields compute-efficient persistence, because the highest-level decomposition is re-invoked only under major context shifts, so long-term steering remains effective across many cycles without continuous user intervention.

\paragraph{Reactive steering.} Users can also issue short, situational commands, such as buying a commodity, sleeping for a period, or performing a work-related action. These commands are operationally different from long-term goals. They are narrow in scope, have clear success criteria, and are intended to take effect immediately. Accordingly, the agent triggers a lightweight planning route, invoking a localized planner to translate the command into an executable action sequence and the Action Simulator to validate feasibility and perform fast repair where needed. This choice reflects a systems principle, that short-horizon clarity should not pay the cost of full-horizon deliberation.

Despite bypassing full top-down re-planning, temporary commands are not one-off overrides that disappear from cognition. Their effects are propagated through the memory system, allowing local interventions to influence future behavior in a principled way. The command's plan, execution result, and outcome are written into $\mathrm{STM}_a(t)$, immediately shaping subsequent planning cycles, and relevant command-associated experiences can consolidate into $\mathrm{LTM}_a(t)$ over time, where they may alter enduring profile items such as habits, preferences, and value tendencies, thereby eventually biasing the highest layers of hierarchical planning. This memory-mediated propagation ensures behavioral continuity, since the agent incorporates the consequences of intervention into its ongoing decision process instead of oscillating between unrelated directives, and it supports personalization through interaction, since repeated steering gradually becomes part of the agent's identity evolution.

\paragraph{Free-form chat.} The third channel is social. Free-form chat does not invoke the planner as an executable command. It enters the dialogue and social-reflection pipeline instead, where it may update social memory, diary context, and later portrait fields. This channel allows users to shape how an agent interprets its experience without turning every utterance into an immediate platform action.

Together these three modes operationalize human-guided and self-directed autonomy. Long-term goals steer what the agent is becoming, temporary commands steer what it should do right now, and chat shapes how it interprets what happens, with all three integrated into a unified loop through hierarchical planning and dual-process memory. All three are bound by the same platform rules as autonomous behavior, so a user can steer an agent toward education, production, recovery, work, or social engagement, but the agent still requires sufficient resources, eligibility, time, and physiological capacity.

\section{Artificial Society }\label{sec:environment}

\subsection{Design Principles}\label{subsec:environment-principles}

\begin{figure}[t]
\centering
\includegraphics[width=\textwidth]{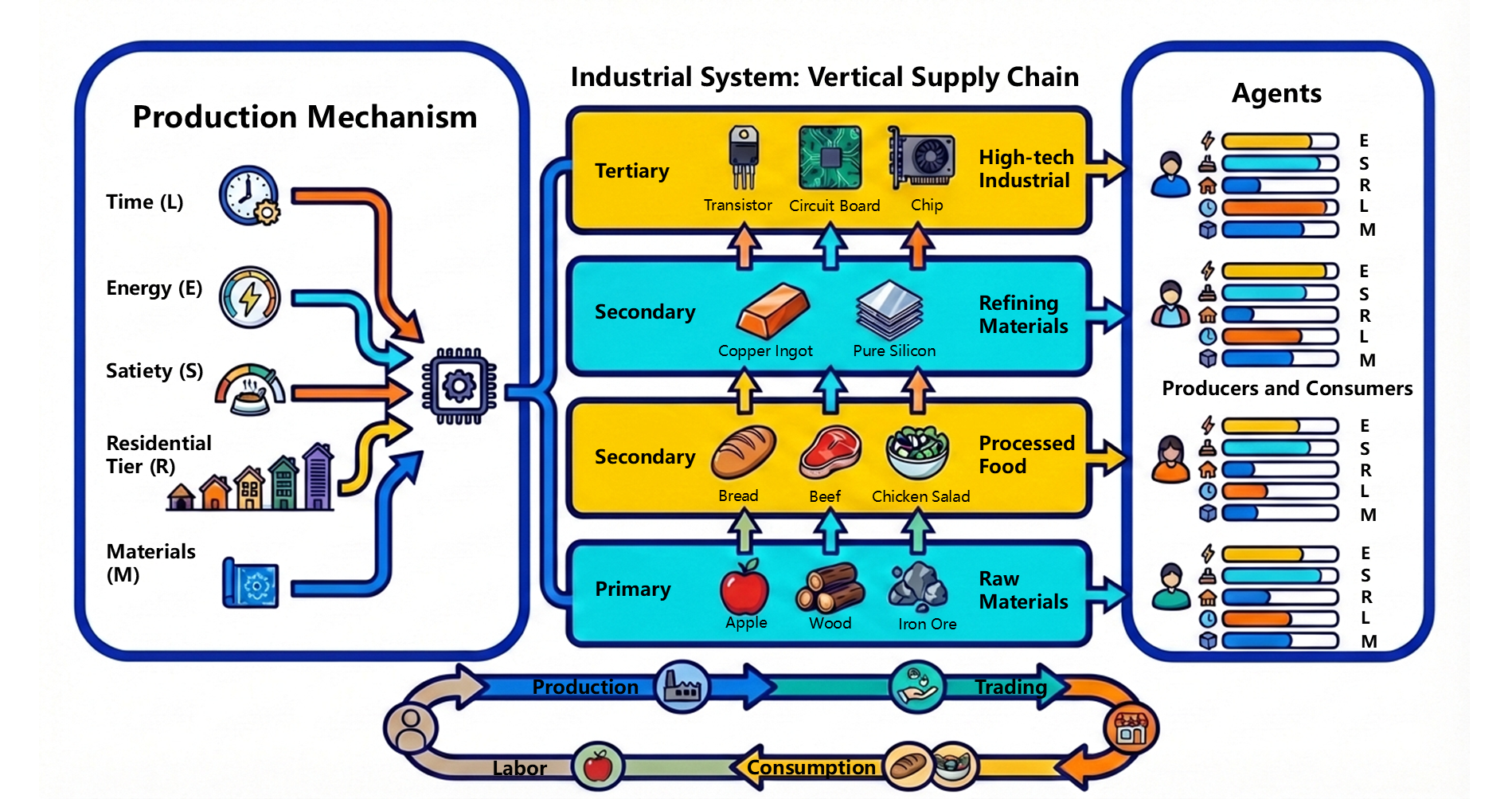}
\caption{Economic environment and production--consumption structure of the simulated world. Production is governed by a resource-constrained production function combining time, energy, satiety, residential tier, and materials. Goods flow through a vertically structured supply chain spanning primary resources, consumer goods, intermediate inputs, and high-technology sectors. Agents act simultaneously as producers and consumers, participating in labor, production, trading, and consumption cycles that jointly determine market dynamics and resource allocation.}
\label{fig:economy-system}
\end{figure}

To support large-scale, long-horizon simulation, the artificial society couples economic production, labor allocation, education, and market exchange into a unified operational framework. Conventional agent-based simulations often treat economic behavior as simplified rule modules or isolated subsystems. Our platform emphasizes structural realism and system-level interdependence. At the core of the design, automated market mechanisms and hierarchical industrial supply chains interact with a dynamic labor market gated by human-capital requirements. In the bottom-up tradition of artificial societies and agent-based computational economics, this closed-loop environment, which emulates real-world labor--production--trade--consumption cycles (Figure~\ref{fig:economy-system}), lets macro-level patterns, such as industrial specialization, wealth stratification, and resource bottlenecks, emerge from micro-level interactions governed by endogenous constraints, without presupposing equilibrium conditions \cite{epstein1996growing,bonabeau2002agent,tesfatsion2002agent}.

By integrating structural constraints, through tiered eligibility and fixed recipes, the design sustains stable long-horizon simulation while preventing degenerate loops, and technological progress emerges organically from cumulative, resource-bound production.
Under these mechanisms, the platform action space is a state-dependent set of affordances. An action is available only when the agent's condition and the environment jointly satisfy its preconditions \cite{simon1955behavioral,gibson2014ecological}. Since those preconditions draw on state that the agent's own actions and other agents' trades continually reshape, executability is a moving target, and a plan that was sound when formed can become infeasible by the time it executes. The environment thereby recreates the strategy--reactivity tension that the agent architecture is built to resolve, and functions as a structured stress test for long-horizon autonomy.

The rest of this section describes the structural mechanisms and defers the default catalogs of commodities, recipes, actions, tiers, and occupations to Appendix~\ref{app:environment-catalogs}.

\subsection{Physiological Constraints and Action Costs}\label{subsec:physiology-action-costs}

Each agent carries three physiological state variables, energy $E_a(t)$, satiety $S_a(t)$, and health $J_a(t)$, whose upper bounds are set by the current residential tier $R_a(t)$. These variables impose homeostatic constraints on reward-seeking behavior, so agents cannot treat production, work, and trade as cost-free symbolic actions \cite{keramati2014homeostatic}. All work and production activities cost satiety and energy. Satiety is replenished through food consumption, with higher-level commodities yielding faster restoration. Energy recovers primarily through sleep, whose hourly recovery rate also depends on residential tier. Health deteriorates through stochastic illness events and cumulative sleep deprivation, and agents can visit the hospital to recover health value. Successful recovery actions are capped by the current residential upper bound. The recovery parameters of individual actions are catalogued in Appendix~\ref{app:environment-catalogs}.

By default, agents start at the full capacity of the first residential tier. Controlled experiments may override the initial values, as specified in Section~\ref{sec:evaluation-protocol}. The platform rejects any action whose post-action energy, satiety, health, balance, or required inventory would become negative, and failed actions consume no time or resources. 

To further internalize sustainability constraints, productive efficiency is modulated by
\begin{equation*}
\mathrm{Eff}_a(t)=G\big(S_a(t),E_a(t),J_a(t),R_a(t),H_a(t)\big),
\label{eq:efficiency}
\end{equation*}
where $G\in(0,1]$ is non-decreasing in each argument and $H_a(t)$ is the education score detailed in Section~\ref{subsec:education-occupation}. This design embeds biologically grounded constraints directly into economic rationality. By making productivity dependent on physiological state, the system penalizes overwork and recovery deprivation through immediate efficiency losses, which creates non-trivial trade-offs between short-term output and long-term capacity, generates organic demand for corresponding time and resources, and ensures that rational agents must strategically balance labor, consumption, and recovery.

The action space supporting these state transitions is summarized in Appendix~\ref{app:environment-catalogs}. It includes education actions, health recovery, energy recovery, satiety recovery, work, commodity production, and market trade. 
\subsection{Production and Supply-Chain Graph}\label{subsec:production-supply-chain}

To emulate the depth of real industrial systems, commodities are organized into a vertical supply-chain taxonomy that induces endogenous division of labor and upstream--downstream dependency \cite{fisher1939production,youno1928increasing}. The primary sector supplies raw materials and basic consumption goods, analogous to agriculture and extractive industries. These goods serve both as essential energy sources for survival and as upstream inputs for industrial production. The secondary sector converts agricultural and mineral resources into high-energy foods and refined industrial intermediates. The tertiary sector comprises high-technology manufacturing, which requires longer production cycles and depends heavily on midstream products. 

Production action is modeled as a conditional synthesis process with hard access constraints and non-substitutable input requirements. For commodity $i$, the platform uses a Leontief-style fixed-proportions rule \cite{leontief1936quantitative}:
\begin{equation}
Y_{a,i}(t)
=
\mathbb{I}\!\left(R_a(t)\ge R^{i}_{\min}\right)
\min\!\left(
\min_{m\in\mathcal{M}_i}\frac{M_{a,m}(t)}{\alpha_{i,m}},
\frac{E_a(t)}{\epsilon_i},
\frac{S_a(t)}{\sigma_i},
\frac{L_a(t)}{\tau_i}
\right),
\label{eq:production}
\end{equation}
where $R^{i}_{\min}$ is the minimum residential tier required to produce commodity $i$, $\mathcal{M}_i$ is the set of required material inputs, $M_{a,m}(t)$ is the amount of input $m$ held by agent $a$, $\alpha_{i,m}$ is the per-unit input requirement, $L_a(t)$ is available labor time, and $\epsilon_i$, $\sigma_i$, and $\tau_i$ are the per-unit energy, satiety, and time costs. All coefficients are exogenously specified recipe parameters and are held fixed within an experiment. The complete commodity catalog and per-unit recipes are given in Appendix~\ref{app:environment-catalogs}.

This design embodies several principles. The recipes form a hierarchical production chain from primary resource extraction to complex high tier products, so disruptions in basic commodity supply propagate throughout the economic system, mirroring real world supply chain dynamics \cite{acemoglu2012network,carvalho2014micro}. Resource costs in energy and satiety generally increase with product complexity, reflecting higher value addition at advanced stages and creating a natural economic gradient. The minimum operator enforces strict non-substitutability, since a shortage in any single factor directly constrains eligibility. Combined with the residential tier barrier that reserves advanced production for agents meeting infrastructure thresholds, our platform prevents unrealistic production surges and ensures that resource scarcities and technological prerequisites propagate coherently across sectors. Advancement through industrial tiers demands cumulative investment in infrastructure, materials, and time, making technological upgrading a resource-bound, path-dependent process \cite{david1985clio,arthur1989competing}.

Certain high-value manufacturing recipes additionally carry a low-probability chance of producing a valuable byproduct. Such lottery-like component introduces behavioral uncertainty into occupational selection and production strategy. In the deployed catalog the byproduct is the Gold Apple, a fixed-value reward item that cannot be targeted as regular output or purchased through the AMM but can be sold and counted in wealth metrics.

\subsection{AMM Market and Price Mechanism}\label{subsec:amm-market}

The commodity system employs a hybrid architecture that functions as both a market exchange and a macro-economic monitoring tool. The core of the exchange is a liquidity-based automated market maker that also serves as an algorithmic central bank creating an elastic money supply \cite{adams2020uniswap,sams2015note}. New currency is minted and injected into circulation when agents sell commodities to a liquidity pool, and currency is withdrawn from circulation upon purchase, which structurally couples the aggregate money supply with real economic output, so that rising productivity and the resulting selling pressure produce a secular expansion of liquidity that reflects genuine economic growth.

At the micro level, each tradable commodity $i$ is paired with the platform currency in a dedicated pool through an inventory supply $\mathrm{IS}_i(t)$ and a currency reserve $\mathrm{CR}_i(t)$. Valuation follows constant-function market-maker designs, using a constant-product relation as the executable pricing rule \cite{adams2020uniswap,angeris2020improved,angeris2021analysis}:
\begin{equation}
\mathrm{IS}_i(t)\mathrm{CR}_i(t)=k_i,
\label{eq:amm-constant-product}
\end{equation}
where $k_i$ is fixed within an experiment for commodity $i$. The public marginal price is
\begin{equation}
p_i(t)=\frac{\mathrm{CR}_i(t)}{\mathrm{IS}_i(t)}.
\label{eq:amm-price}
\end{equation}
Finite trades experience price impact and slippage, which is a standard concern in market microstructure and AMM analysis \cite{kyle1985continuous,angeris2021analysis}. In our platform, trades carry no transaction fee and agents may submit arbitrary positive quantities. If an agent buys $b$ units of commodity $i$, the pool inventory falls to $\mathrm{IS}_i(t)-b$ and the currency paid to the pool is
\begin{equation}
\Delta \mathrm{CR}^{\mathrm{buy}}_i(b)
=
\frac{k_i}{\mathrm{IS}_i(t)-b}-\mathrm{CR}_i(t),
\qquad
0<b<\mathrm{IS}_i(t).
\label{eq:amm-buy}
\end{equation}
If an agent sells $s$ units, the pool inventory rises to $\mathrm{IS}_i(t)+s$ and the currency received from the pool is
\begin{equation}
\Delta \mathrm{CR}^{\mathrm{sell}}_i(s)
=
\mathrm{CR}_i(t)-\frac{k_i}{\mathrm{IS}_i(t)+s},
\qquad
s>0.
\label{eq:amm-sell}
\end{equation}
The corresponding effective prices, $\Delta \mathrm{CR}^{\mathrm{buy}}_i(b)/b$ and $\Delta \mathrm{CR}^{\mathrm{sell}}_i(s)/s$, generally differ from the initial marginal price in Equation~\eqref{eq:amm-price}. As a commodity is purchased from the pool, $\mathrm{IS}_i(t)$ falls, $\mathrm{CR}_i(t)$ rises, and the marginal price increases continuously, signaling growing scarcity while selling has the opposite effect. All price signals are therefore direct, real-time reflections of agent-driven supply and demand.

While the AMM governs real-time prices, the system simultaneously computes a macro-level price trend index to monitor economy-wide inflation, in the spirit of elementary price-index construction \cite{diewert1995axiomatic}. For commodity $i$, with initial baseline price $p_i(0)$ and current price $p_i(t)$, the price change ratio is
\begin{equation}
\mathrm{PCR}_i(t)=\frac{p_i(t)}{p_i(0)}.
\label{eq:pcr}
\end{equation}
To prevent extreme price spikes of a single item from distorting the aggregate measure, category indices for food and non-food commodities are computed as geometric means of the corresponding ratios:
\begin{align}
\overline{\mathrm{PCR}}_{\mathrm{food}}(t)
&=
\left(\prod_{i=1}^{n_{\mathrm{food}}}\mathrm{PCR}_i(t)\right)^{1/n_{\mathrm{food}}},
\nonumber\\
\overline{\mathrm{PCR}}_{\mathrm{nonfood}}(t)
&=
\left(\prod_{j=1}^{n_{\mathrm{nonfood}}}\mathrm{PCR}_j(t)\right)^{1/n_{\mathrm{nonfood}}}.
\label{eq:pcr-category}
\end{align}
The overall price index is then
\begin{equation}
\overline{\mathrm{PCR}}_{\mathrm{overall}}(t)
=
\frac{n_{\mathrm{food}}}{N}\overline{\mathrm{PCR}}_{\mathrm{food}}(t)
+
\frac{n_{\mathrm{nonfood}}}{N}\overline{\mathrm{PCR}}_{\mathrm{nonfood}}(t),
\label{eq:pcr-overall}
\end{equation}
where $N=n_{\mathrm{food}}+n_{\mathrm{nonfood}}$. The index does not alter transaction prices. It serves as an input to other economic modules, in particular the nominal wage rules of Section~\ref{subsec:education-occupation} (Equations~\eqref{eq:wage-static}--\eqref{eq:wage-dynamic}), and as a platform-level signal of the economy's inflationary or deflationary state. Food commodities are the edible commodities listed in Table~\ref{tab:food-recovery}.

This dual-component design strengthens the socio-economic coherence of the artificial economy. The AMM layer gives every transaction an immediate and localized market impact, creating an authentic feedback loop between individual actions and resource availability and anchoring valuation to agent-generated activity. The index layer allows the platform to observe and respond to emergent society-scale trends, coupling micro-level behavior with macro-level stability, as nominal wages adjustment drives feed back into agents' purchasing power and trading decisions, which move those same prices. So price dynamics both reflect and shape the evolution of the artificial economy.

\subsection{Education-Occupation Gating}\label{subsec:education-occupation}
Education in our platform means agents spend time and resources on study actions to gain education scores \cite{becker1962investment,mincer1974schooling}. Education score $H_a(t)$ accumulates through unit study actions:
\begin{equation}
H_a(t+\tau_u)=H_a(t)+g_u,
\label{eq:education}
\end{equation}
where $u$ is the study unit, $\tau_u$ is its game-time duration, and $g_u$ is the education increment attached to that unit. In the default action catalog, paid learning, reading, and self-study each increase education by one unit, but they differ in time and resource cost. The education score serves two purposes, enhancing productive efficiency through Equation~\eqref{eq:efficiency} and acting as a credential for occupational access. Agents therefore face a strategic trade-off, foregoing immediate labor income and consumption to accrue long-run competitive advantage and balance educational investment against immediate survival needs.

Occupations are organized into six tiers. Each occupation $j$ has a minimum residential tier requirement $R^{(j)}_{\min}$, a knowledge floor $H^{(j)}_{\mathrm{floor}}$, an eligibility-share parameter $\pi_j\in(0,1]$, a base wage $w^{(j)}_0$, and, for some tiers, a prerequisite commodity held in inventory. The education requirement is population-relative, admitting approximately the top $\pi_j$ fraction of agents by education score. Let $F_t$ be the empirical distribution of education scores over agents active at time $t$; the effective knowledge threshold is
\begin{equation}
\widehat{H}^{(j)}_{\min}(t)=\max\!\Big(H^{(j)}_{\mathrm{floor}},\ q_{1-\pi_j}(t)\Big),
\label{eq:dynamic-threshold}
\end{equation}
where $q_{1-\pi_j}(t)=\inf\{h:F_t(h)\ge 1-\pi_j\}$ is the $(1-\pi_j)$ quantile of the current education distribution. As the population's average education rises, so do the effective requirements for high-tier jobs, preventing credential devaluation and preserving incentives for continuous human-capital investment \cite{collins2019credential,berg1970education,frank2012darwin}. Moreover, our platform anchors eligibility to a commodity and residential tier hard floor. Let $P_{a,j}(t)=1$ indicate that the prerequisite commodity condition for occupation $j$ is met, or the requirement is none. Agent $a$ is eligible to apply for occupation $j$ if
\begin{equation}
\mathbb{I}\!\left(H_a(t)\ge \widehat{H}^{(j)}_{\min}(t)\right)
\mathbb{I}\!\left(R_a(t)\ge R^{(j)}_{\min}\right)
\mathbb{I}\!\left(P_{a,j}(t)=1\!\right)=1.
\label{eq:occupation-eligibility}
\end{equation}
Occupation access is thus conditional on education relative to the current population, residential progress, and commodity accumulation, These rules create a skill and access structured labor market \cite{autor2003skill}, which reflect two foundational dimensions of real-world inequality, human-capital investment operationalized through the education score, and asset-based access proxied by residential tier \cite{schultz1961investment,becker1962investment}. Only after satisfying both the educational and the capital requirement threshold can an agent qualify for higher-tier occupations, so upward mobility remains contingent on sustained investment in both material and human capital.

Eligibility does not by itself imply a job transition. Recruitment proceeds in a cycle of fixed simulation time. At the start of each cycle the available job list is refreshed. Each agent decides for itself whether to apply, and at most one application per agent takes effect in a given cycle. Applications satisfying the eligibility rule of Equation~\eqref{eq:occupation-eligibility} are reviewed by an AI-driven mayor agent, which evaluates the applicants for each position and selects suitable candidates from the content of their applications. Agents not selected remain unemployed for that cycle.

The wage structure is bifurcated into a static and a dynamic regime, both modulated by the overall price index of Equation~\eqref{eq:pcr-overall} so that nominal wages reflect changes in the aggregate price level. The static regime applies to lower-tier occupations, whose wage scales a predetermined base value by the price index,
\begin{equation}
w^{(j)}(t)=w^{(j)}_0\,\overline{\mathrm{PCR}}_{\mathrm{overall}}(t),
\label{eq:wage-static}
\end{equation}
ensuring a stable and predictable income floor for foundational roles, insulated from the competitive pressures of higher-tier labor markets. The dynamic regime governs higher-tier occupations, whose wages respond also to the competitive landscape through the effective knowledge threshold,
\begin{equation}
w^{(j)}(t)=w^{(j)}_0\,\Phi\!\big(\widehat{H}^{(j)}_{\min}(t)\big)\,\overline{\mathrm{PCR}}_{\mathrm{overall}}(t)\,(1+\delta_t),
\label{eq:wage-dynamic}
\end{equation}
where $\Phi(\cdot)$ is a non-decreasing function translating a rising knowledge threshold into a wage premium, so that higher entry barriers command greater compensation \cite{becker1962investment,mincer1974schooling}, and $\delta_t\in[-\bar\delta,\bar\delta]$ is a bounded short-term adjustment reflecting temporary labor-market shocks. This dual-regime structure mirrors real world labor market segmentation \cite{doeringer2020internal}, keeping compensation for competitive occupations aligned with evolving societal capabilities while the static component anchors expectations and guarantees a minimum livelihood. The tier catalog in Appendix~\ref{app:environment-catalogs} marks which occupations follow each regime. 

Participation in any occupation incurs per-hour physiological costs. To keep working, agents must earn wages sufficient to afford the consumables that restore their capacity to work. This linkage places a subsistence constraint on labor supply, determined jointly by nominal wages, commodity prices, and the physiological demands of occupations. Education and residential investment can improve future access while draining short-term resources, so survival, delayed human-capital investment, and production planning compete for the same limited budget.

\section{Evaluation}\label{sec:evaluation-protocol}

The evaluation addresses five questions. On the environment side, we examine whether the deployed society sustains stable, non-degenerate activity over months of public operation, and whether our designs give rise to market and stratification dynamics of the kind documented for real economies. On the agent side, we quantify the contribution of hierarchical branching, objective decomposition, and pre-execution simulation when agents pursue interacting objectives, test whether successive portrait updates form coherent identity trajectories over long horizons, and estimate the association between human steering and that evolution.

Two classes of evidence support these questions. Controlled small-batch experiments compare architecture variants under identical task and initialization conditions and support component-level conclusions about the architecture. Public-deployment observation records the behavior of the platform after its public launch in late August 2025, during which human users created, instructed, and conversed with autonomous agents in a persistent environment. It supports deployment-scope observational claims about the society and its agents. The following subsections present each experiment, giving its design, data, metrics, and results together. Remaining aggregation conventions and full model specifications appear in Appendix~\ref{app:evaluation-details}.

\subsection{Public-Deployment Economy Analyses}\label{subsec:deployment-economy-protocol}

The economy analyses evaluate whether the deployed platform sustains stable executable activity and whether our designs generate non-trivial market and stratification dynamics.

\subsubsection{Market Stability and Stylized Facts}\label{subsec:market-experiment}

We use transaction price of commodities for the following analyses. Prices are aggregated into five-minute open-high-low-close intervals, the last traded price in each interval serves as the close price, and log returns $r_{i,t}=\log P_{i,t}-\log P_{i,t-1}$ are computed from the close prices $P_{i,t}$ of commodity $i$, with $V_{i,t}$ its traded volume. The reward item Gold Apple is excluded from the diagnostics.

Market behavior is summarized by two groups of indicators. Stability indicators, the log-price range and the maximum drawdown, check that the economy neither diverges nor freezes. A stable market should combine a small range and drawdown with persistent, non-degenerate fluctuation. Stylized-fact indicators, covering heavy tails, return asymmetry, volatility clustering and its statistical significance, and volume--volatility co-movement, evaluate whether returns reproduce the regularities documented for real financial markets \cite{cont2001empirical,mandelbrot1963variation,engle1982autoregressive,ljung1978measure}. Full definition of the indicators can be found in Appendix~\ref{app:economy-details}.

\begin{figure}[t]
\centering
\includegraphics[width=\linewidth]{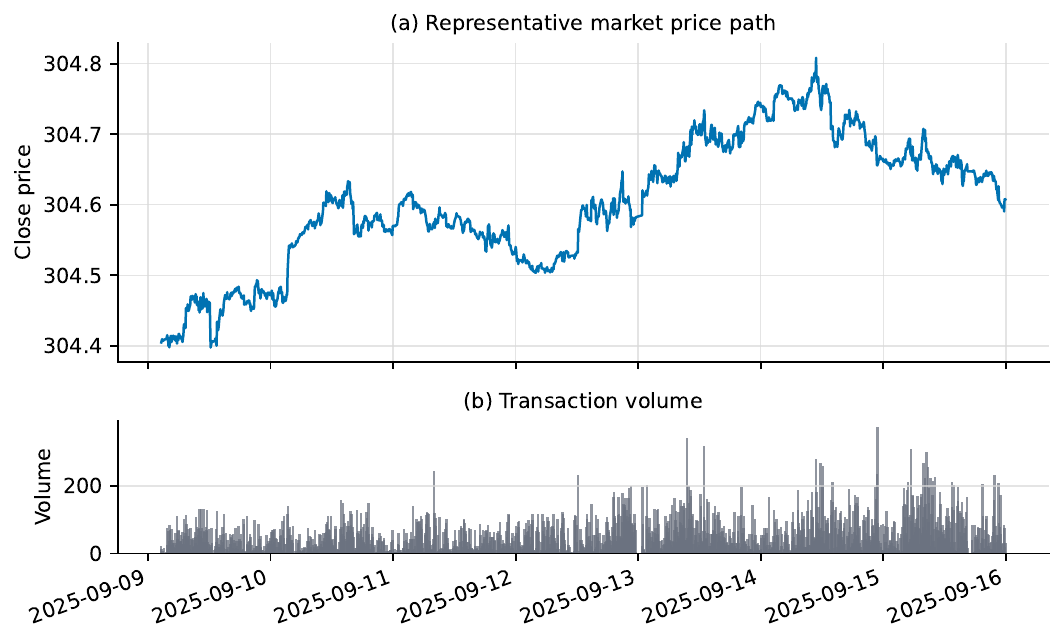}
\caption{Representative market trace from the public deployment, showing the fish market from 2025-09-09 to 2025-09-15 in real-world time. The top panel plots the five-minute close price and the bottom panel the corresponding transaction volume. The close price stays within a log-price range of 0.00135 with a maximum drawdown of 0.0715\%, while price and volume fluctuation persists throughout the window.}
\label{fig:fish-market}
\end{figure}

The market analysis covers 555{,}402 five-minute OHLCV bars across 23 item series from the mature deployment phase, excluding the Gold Apple reward item. The transaction-derived logs form continuous price series, with no degenerate constant prices and no explosive numerical behavior. Figure~\ref{fig:fish-market} shows a representative fish-market window from 2025-09-09 to 2025-09-15 in real-world time. Across 1{,}815 five-minute bars, the close price remains bounded between 304.398 and 304.808, corresponding to a log-price range of 0.00135; the maximum drawdown is 0.0715\%. The AMM-based economy remained numerically stable in this representative window while still generating persistent micro-fluctuations and transaction volume during deployment.

\begin{figure}[t]
\centering
\includegraphics[width=\textwidth]{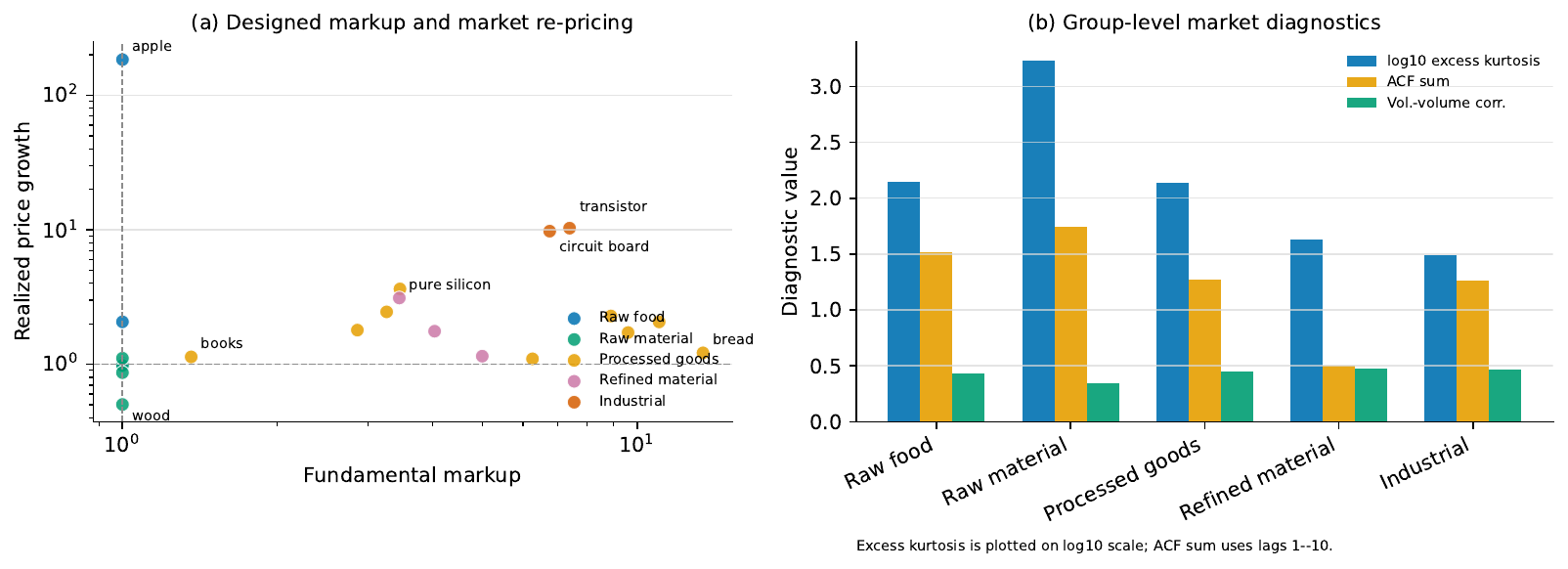}
\caption{Market structure diagnostics in the public deployment. Panel A relates designed fundamental markup to realized price growth. Panel B summarizes group-level stylized-fact diagnostics by supply-chain position; excess kurtosis is plotted on a $\log_{10}$ scale, while the volatility-memory score sums the first ten autocorrelations of absolute returns.}
\label{fig:environment-market-structure}
\end{figure}

The same price series also reproduce the stylized facts targeted by the second indicator group. All 23 market-pool commodities show positive excess kurtosis, positive lag-1 autocorrelation of absolute returns, and positive volume--volatility correlation, with cross-commodity medians of 137.0, 0.186, and 0.436. Ljung--Box tests over the first ten lags reject the absence of serial dependence in absolute returns for every commodity (all $p<10^{-8}$), suggesting that the volatility clustering is statistically significant throughout the market system. Returns in the deployed economy are thus heavy-tailed and volatility-clustered, far from independent price noise. Supply-chain position adds a descriptive layer to this picture. Industrial commodities show the largest median realized growth (10.03), while raw materials show the largest median excess kurtosis (1708.0; Figure~\ref{fig:environment-market-structure}b and Appendix Table~\ref{tab:app-environment-group-diagnostics}). So price drift concentrates downstream in the production chain while tail risk concentrates upstream.

This growth pattern raises the question of whether realized prices simply reproduce the price structure fixed at design time. To examine this, we compare each commodity's realized growth with its fundamental markup, which equals to the initial price $p_i(0)$ divided by the raw-material cost implied by the commodity's recipe chain. It is a designed price--cost ratio in the sense of classical and modern markup measures. Commodities near the unit-growth line trade at their designed values and departures indicate endogenous repricing \cite{lerner1934concept,de2020rise}. Figure~\ref{fig:environment-market-structure}a plots realized growth against this markup on logarithmic axes. Realized growth is not a mechanical function of the markup. The industrial chain appreciates roughly tenfold under sustained production demand, consumption demand lifts apple far above its unit markup, wood falls below parity under abundant supply, and high-markup processed foods appreciate only modestly in the absence of comparable demand. Prices in the deployed economy are thus discovered through trading under the coupled demand structure of the society, with the designed price ladder serving as an anchort.

\begin{table}[t]
\caption{Stylized-fact diagnostics for representative public-deployment commodity markets, computed from five-minute log returns. Growth is realized growth $P_T/P_0$; ACF$_1(|r_t|)$ is the lag-1 autocorrelation of absolute returns; Vol.--volume corr.\ is $\mathrm{corr}(|r_t|,\log(1+V_t))$. Indicator definitions are given in Appendix~\ref{app:economy-details}.}
\label{tab:volatility-analysis}
\centering
\small
\begin{tabular}{@{}lccccc@{}}
\toprule
Commodity & Growth & Excess kurtosis & Skewness & ACF$_1(|r_t|)$ & Vol.--volume corr. \\
\midrule
Apple & 184.39 & 34.5 & $-0.47$ & 0.401 & 0.527 \\
Wood & 0.50 & 2962.9 & $-43.44$ & 0.577 & 0.167 \\
Fish & 1.01 & 44.6 & $-0.72$ & 0.227 & 0.426 \\
Books & 1.13 & 1549.1 & $-10.10$ & 0.205 & 0.097 \\
Copper ingot & 1.15 & 38.1 & 1.08 & 0.140 & 0.425 \\
Pure silicon & 3.11 & 47.4 & 0.23 & 0.107 & 0.475 \\
Transistor & 10.28 & 27.0 & $-1.72$ & 0.186 & 0.458 \\
Circuit board & 9.77 & 34.6 & $-2.15$ & 0.197 & 0.469 \\
\bottomrule
\end{tabular}
\end{table}

\subsubsection{Education--Wealth Stratification}\label{subsec:stratification-experiment}

A socioeconomic simulation should generate structured inequality through the interactions of its agents, and observed stratification is therefore a standard test of the deployed society \cite{schelling1971dynamic,epstein1996growing}. We quantify the education--wealth gradient among employed characters with the quadratic model
\begin{equation}
\log(1+W_a)=\alpha+\beta_1 H_a+\beta_2 H_a^2+\varepsilon_a,
\label{eq:stratification-model}
\end{equation}
where net worth $W_a=B_a+\sum_{i\in\mathcal{I}_{\mathrm{amm}}}p_i q_{a,i}+\sum_{r\in\mathcal{I}_{\mathrm{fix}}}v_r q_{a,r}$ sums the currency balance $B_a$, the held quantities $q_{a,i}$ of AMM-priced market commodities $i\in\mathcal{I}_{\mathrm{amm}}$, and the fixed-value reward items $r\in\mathcal{I}_{\mathrm{fix}}$ valued at $v_r$, and $H_a$ is the education score. The sample is restricted to 11{,}856 employed characters because wages are the channel through which education-gated occupations generate income. Coefficients are estimated by ordinary least squares with HC3 heteroskedasticity-robust standard errors.

\begin{figure}[t]
\centering
\includegraphics[width=\textwidth]{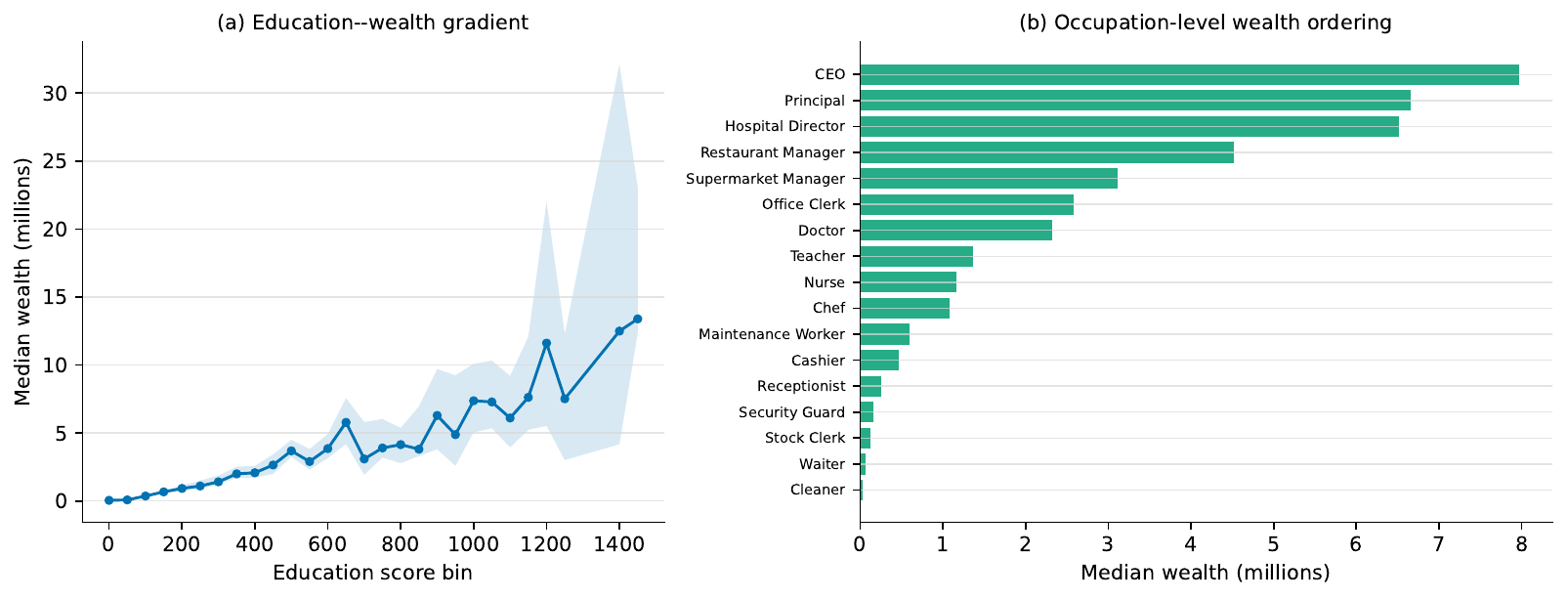}
\caption{Education and occupation stratification in the public deployment. Panel A shows median wealth by education-score bin, using width-50 bins over scores in $[0,1500)$ and retaining bins with at least 20 employed characters; shaded bands are percentile-bootstrap 95\% confidence intervals for the bin medians. Panel B reports occupation-level median wealth, excluding unemployed characters.}
\label{fig:environment-stratification}
\end{figure}

Education and wealth are positively associated (Spearman $\rho=0.648$, $p<10^{-8}$), and the model explains a third of the variance in log wealth (adjusted $R^2=0.327$). The gradient is approximately log-linear over the observed range. Each additional education point is associated with a 0.45\% increase in net worth ($\beta_1=0.00453$, $p<10^{-8}$), while the quadratic term is small and not significant ($\beta_2=-3.8\times10^{-7}$, $p=0.079$). The strong rank correlation together with the moderate $R^2$ indicates a robustly monotone yet individually noisy relationship, as expected under the heavy-tailed wealth distribution. Median wealth rises from about 0.05 million coins in the lowest education bin to 13.39 million coins in the 1450 education-score bin (Figure~\ref{fig:environment-stratification}a), and occupation-level medians range from 0.026 million coins for Cleaners to 7.97 million coins for CEOs (Figure~\ref{fig:environment-stratification}b and Appendix Table~\ref{tab:app-occupation-sorting}). Notice that the existence of this gradient is designed, since education, residential tier, and occupation access are explicitly coupled (Section~\ref{subsec:education-occupation}). What emerges in deployment is the sorting process, through which the population differentiates into persistent strata under the dynamic thresholds and wage regimes of the labor market.

\subsection{Controlled Experiments on the Agent Architecture}\label{subsec:controlled-protocol}

We conduct controlled experiments on the agent architecture to attribute its capability to individual components. The ablation removes branch decomposition and objective decomposition in turn, measuring each component's contribution to task outcomes across four task families. The trace audit works at the mechanism level, pairing each planner-emitted candidate sequence with its simulator-approved counterpart to quantify how pre-execution simulation intervenes in, repairs, and preserves the planner's intent.

\subsubsection{Controlled Ablation}

The ablation isolates the contributions of hierarchical branching and objective decomposition by comparing three planner variants under identical environmental conditions. The full architecture runs the complete hierarchical planning structure of Section~\ref{sec:architecture}. \textit{Without-Branch} removes branch decomposition and restricts planning to a single reasoning branch. \textit{Without-OD} removes structured objective decomposition, so parallel branches generate action lists without intermediate objective factorization.

Each variant is instantiated with 80 agents. Profiles are initialized uniformly except for the personality field, where the 16 MBTI-style seed categories are represented by five agents each. Initial health, satiety, and energy are set to 60; balance, education score, and inventory are reset to zero. The controlled runs use a $35\times$ game-time acceleration, with all time-dependent processes scaled consistently.

The variants are evaluated on four task families, denoted E1--E4 (Table~\ref{tab:controlled-experiment-codes}); each task is issued to the agent as its top-level long-horizon objective. E1 combines high-technology production with income generation and physiological maintenance, so that crafting, earning, and state upkeep compete for the same resources over a long horizon. E2 instructs agents to accumulate wealth and educational experience simultaneously, pairing immediate income with a delayed-return investment. E3 is deliberately specified in vague and abstract terms, asking agents to explore diverse actions without enumerating concrete subgoals or success criteria. E4 is a narrow single-objective command to craft chips as efficiently as possible, and is evaluated over the first 16 hours of each controlled trace. E1 and E2 exercise multi-objective coordination, while E3 and E4 probe open-ended exploration and direct execution. Outcomes are task-specific and are reported alongside each task below. The per-task outcome metrics, statistical conventions, and aggregation details are recorded in Appendix~\ref{app:evaluation-details}.

\begin{table}[t]
\caption{Task families used in the controlled experiments. Each task is issued to the agent as its top-level long-horizon objective.}
\label{tab:controlled-experiment-codes}
\centering
\small
\begin{tabular}{@{}lp{0.72\linewidth}@{}}
\toprule
Code & Controlled task family \\
\midrule
E1 & High-tech production with physiological state maintenance \\
E2 & Wealth accumulation with education investment \\
E3 & Action-diversity exploration \\
E4 & Direct efficient chip production \\
\bottomrule
\end{tabular}
\end{table}

\begin{figure}[t]
\centering
\includegraphics[width=\textwidth]{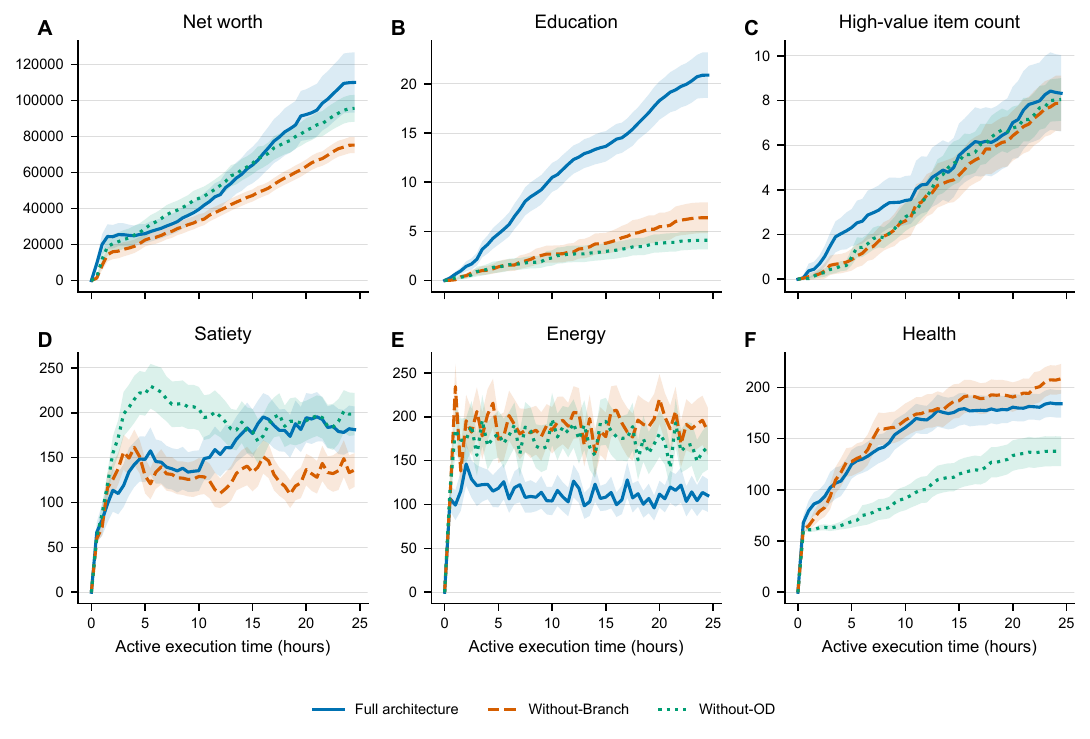}
\caption{Outcome trajectories for E1 ($n=80$ agents per planner variant). Lines show planner-level means aligned on each agent's active execution time; shaded bands are pointwise 95\% confidence intervals over agents. Education diverges progressively under the full architecture, while net worth, high-value holdings, and physiological states remain heterogeneous across variants.}
\label{fig:ablation-complex-trajectory}
\end{figure}

\paragraph{E1: high-tech production with state maintenance.} E1 evaluates whether agents can balance production performance with state upkeep, comparing high-value item output, net worth, education, and terminal physiological state (Table~\ref{tab:ablation-complex-state}). The full architecture attains the highest mean net worth and currency balance. Physiological states remain stable under all three planners. The full architecture ranks second in both satiety and health, with Without-OD leading satiety and Without-Branch leading health. Their lower terminal energy reflects the heavier production workload they sustain. The clearest separation concerns delayed-return learning. The full architecture reaches a mean education score of 20.90, against 6.39 under Without-Branch and 4.10 under Without-OD, a large and statistically supported separation, and Figure~\ref{fig:ablation-complex-trajectory} shows the difference accumulating progressively over the run. The full planner strategically allocates learning actions that raise future production efficiency, coordinating delayed investment with long-term returns while continuing to manage production and physiological constraints.

\begin{table}[t]
\caption{Terminal outcomes for E1, high-tech production with state maintenance ($n=80$ agents per planner variant). Values are mean $\pm$ SD over agents. Net worth combines currency balance and inventory value, and the high-value item count sums terminal holdings of the five highest-priced market commodities in the controlled configuration. Pairwise contrast statistics are given in Appendix Table~\ref{tab:app-complex-contrasts}.}
\label{tab:ablation-complex-state}
\centering
\scriptsize
\setlength{\tabcolsep}{5pt}
\textbf{(a) Task outcomes}\\[2pt]
\begin{tabular}{@{}lrrrr@{}}
\toprule
Planner & Balance & Net worth & Education & High-value items \\
\midrule
Without-Branch & 11{,}573 $\pm$ 15{,}447 & 75{,}292 $\pm$ 20{,}188 & 6.39 $\pm$ 7.13 & 7.88 $\pm$ 5.64 \\
Without-OD & 29{,}946 $\pm$ 34{,}883 & 95{,}714 $\pm$ 34{,}012 & 4.10 $\pm$ 4.25 & 8.05 $\pm$ 4.35 \\
Full architecture & 39{,}097 $\pm$ 58{,}152 & 110{,}098 $\pm$ 76{,}334 & 20.90 $\pm$ 10.59 & 8.33 $\pm$ 7.80 \\
\bottomrule
\end{tabular}

\vspace{4pt}
\textbf{(b) Terminal physiological state}\\[2pt]
\begin{tabular}{@{}lrrr@{}}
\toprule
Planner & Satiety & Energy & Health \\
\midrule
Without-Branch & 136.14 $\pm$ 85.53 & 183.30 $\pm$ 123.28 & 208.45 $\pm$ 65.98 \\
Without-OD & 198.45 $\pm$ 107.68 & 165.58 $\pm$ 117.83 & 137.63 $\pm$ 66.53 \\
Full architecture & 181.14 $\pm$ 117.69 & 110.25 $\pm$ 85.12 & 184.03 $\pm$ 61.31 \\
\bottomrule
\end{tabular}
\end{table}

\begin{figure}[htbp]
\centering
\includegraphics[width=\textwidth]{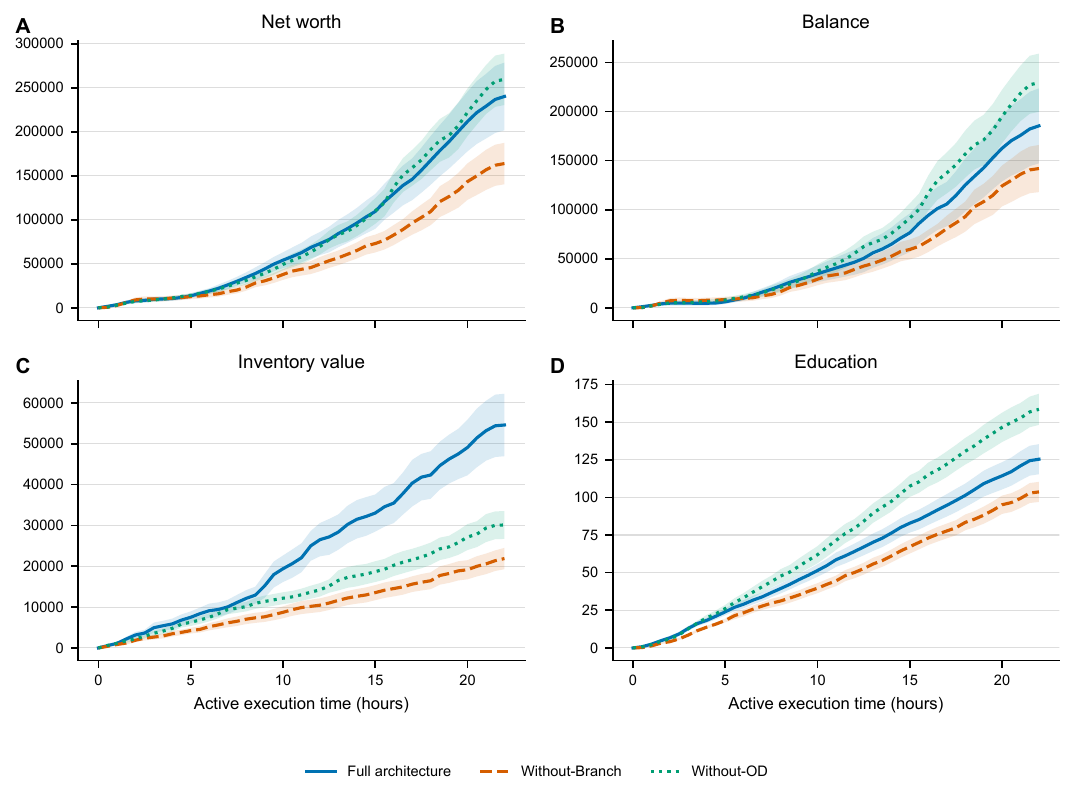}
\caption{Outcome trajectories for E2 ($n=80$ agents per planner variant). Lines show planner-level means aligned on each agent's active execution time; shaded bands are pointwise 95\% confidence intervals over agents. The differences among variants accumulate gradually over the run instead of arising near termination.}
\label{fig:app-ablation-wealth-trajectory}
\end{figure}

\paragraph{E2: wealth and education.} E2 pairs income maximization with education accumulation (Table~\ref{tab:ablation-money-study}). Without-Branch performs worst across metrics, trailing the other two variants in net worth, inventory value, and education. Without-OD slightly outperforms the full architecture on currency balance and education, while the full architecture retains the largest inventory value. Figure~\ref{fig:app-ablation-wealth-trajectory} shows these differences accumulating over the run. Educational activities in this environment involve no prerequisite structure, unlike industrial production with its recipes and upstream material chains. Therefore, objective decomposition offers limited benefit here and its planning overhead can reduce task efficiency. The full architecture does not produce the strongest outcome on every task. Its advantage concentrates where objectives interact through resource, time, and production constraints.

\begin{table}[t]
\caption{Terminal outcomes for E2, wealth accumulation with education investment ($n=80$ agents per planner variant). Values are mean $\pm$ SD over agents; net worth combines currency balance and inventory value. Pairwise contrast statistics are given in Appendix Table~\ref{tab:app-wealth-contrasts}.}
\label{tab:ablation-money-study}
\centering
\scriptsize
\setlength{\tabcolsep}{5pt}
\resizebox{\textwidth}{!}{%
\begin{tabular}{@{}lrrrr@{}}
\toprule
Planner & Balance & Inventory value & Net worth & Education \\
\midrule
Without-Branch & 141{,}955 $\pm$ 109{,}666 & 21{,}888 $\pm$ 11{,}919 & 163{,}843 $\pm$ 107{,}396 & 103.63 $\pm$ 30.48 \\
Without-OD & 229{,}416 $\pm$ 135{,}516 & 30{,}098 $\pm$ 15{,}595 & 259{,}514 $\pm$ 133{,}254 & 158.48 $\pm$ 47.22 \\
Full architecture & 185{,}470 $\pm$ 174{,}387 & 54{,}584 $\pm$ 34{,}899 & 240{,}054 $\pm$ 175{,}430 & 125.34 $\pm$ 45.68 \\
\bottomrule
\end{tabular}
}
\end{table}

\begin{figure}[htbp]
\centering
\includegraphics[width=\textwidth]{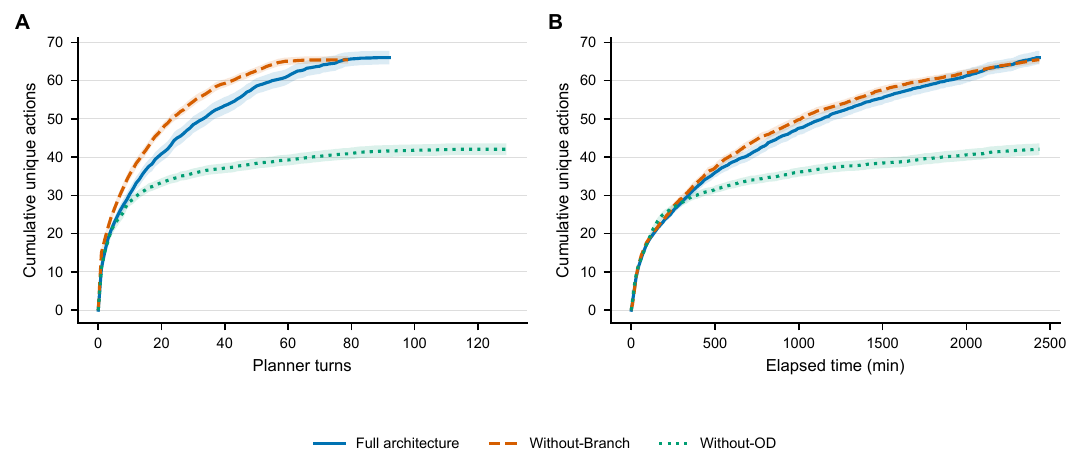}
\caption{Coverage trajectories for E3 ($n=80$ agents per planner variant). Panel A aligns agents by planning turn and Panel B by elapsed wall-clock time. Lines show planner-level means; shaded bands are pointwise 95\% confidence intervals over agents. Without-OD plateaus near 42 unique actions, while the full architecture sustains exploration when coverage is normalized by time.}
\label{fig:app-ablation-diversity-trajectory}
\end{figure}

\paragraph{E3: action diversity.} E3 issues a deliberately vague exploration goal (Table~\ref{tab:ablation-diversity}). Without-OD reaches its exploration boundary at roughly 42 unique actions and stagnates there, well below the comparable totals of the full architecture and Without-Branch (66.01 versus 65.43). The planning-iteration view in Figure~\ref{fig:app-ablation-diversity-trajectory} refines this picture. Without-Branch explores more actions per planning turn and converges to its boundary sooner. Since actions in this environment carry explicit durations, once exploration is normalized by active time the full architecture explores more unique actions per minute (0.033 versus 0.028). The full architecture thus accounts for fine-grained temporal costs and sustains exploration under realistic execution budgets.

\begin{figure}[htbp]
\centering
\includegraphics[width=\textwidth]{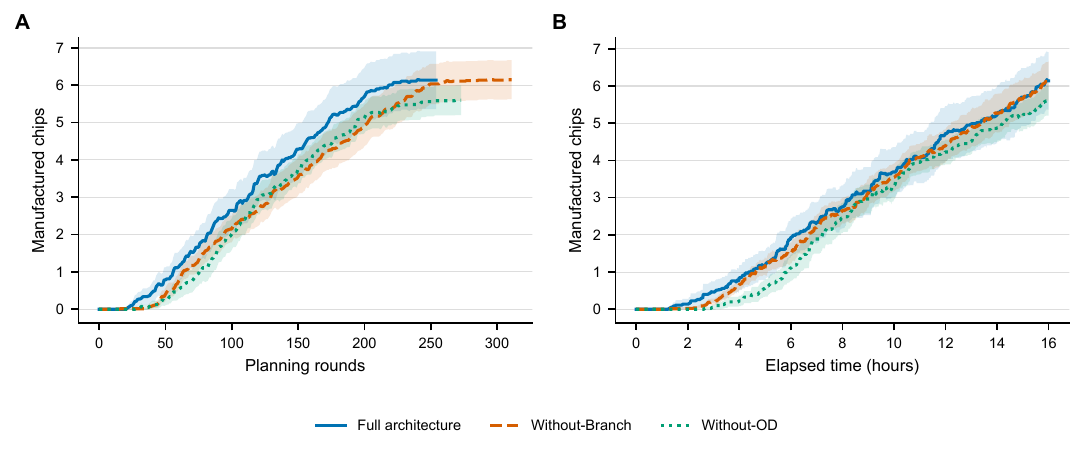}
\caption{Production trajectories for E4 ($n=80$ agents per planner variant). Panels A and B align manufactured-chip counts by planning round and by elapsed time. Lines show planner-level means; shaded bands are pointwise 95\% confidence intervals over agents. The three variants remain closely matched throughout the task.}
\label{fig:app-ablation-chip-trajectory}
\end{figure}

\paragraph{E4: efficient chip production.} E4 represents tasks with a clearly specified goal, a concise procedure, and a direct execution path, the regime targeted by the lightweight planning route of Section~\ref{subsec:human-steering-architecture}. The full architecture and Without-Branch achieve nearly identical manufactured-chip counts (6.14 versus 6.15), Without-OD trails slightly (5.60). Figure~\ref{fig:app-ablation-chip-trajectory} shows the corresponding evolution over planning rounds and elapsed time. For well-defined objectives with minimal structural complexity, lightweight planning is sufficient and effective. This result supports the design of the reactive steering channel, in which a direct human command wakes only part of the architecture, a localized planner together with the pre-execution simulator, and leaves the full branch-thinking stack dormant. The reduced planning overhead improves response efficiency to external instructions without sacrificing task performance.

\begin{table}[t]
\caption{Terminal outcomes for E3, action-diversity exploration, and E4, efficient chip production ($n=80$ agents per planner variant). Values are mean $\pm$ SD over agents. The first three columns report E3 action coverage; the final column reports E4 manufactured-chip counts. Pairwise contrast statistics are given in Appendix Tables~\ref{tab:app-diversity-contrasts} and~\ref{tab:app-chip-contrasts}.}
\label{tab:ablation-diversity}
\centering
\scriptsize
\setlength{\tabcolsep}{3pt}
\resizebox{\textwidth}{!}{%
\begin{tabular}{@{}lrrrr@{}}
\toprule
Planner & Unique actions/turn & Unique actions/min & Total unique actions & Manufactured chips \\
\midrule
Full architecture & 1.11 $\pm$ 1.47 & 0.033 $\pm$ 0.045 & 66.01 $\pm$ 8.15 & 6.14 $\pm$ 3.53 \\
Without-Branch & 1.14 $\pm$ 0.33 & 0.028 $\pm$ 0.006 & 65.43 $\pm$ 4.33 & 6.15 $\pm$ 2.39 \\
Without-OD & 0.73 $\pm$ 1.08 & 0.030 $\pm$ 0.049 & 42.06 $\pm$ 7.06 & 5.60 $\pm$ 1.80 \\
\bottomrule
\end{tabular}
}
\end{table}

\subsubsection{Action Simulator Trace Audit}\label{subsec:audit-experiment}

While the ablation measures component contributions at the outcome level, the trace audit examines the mechanism, quantifying how pre-execution simulation converts planner proposals into executable action sequences and whether the planner's intent survives this repair. The audit pairs each planning episode's candidate action multiset $C$ with the simulator-approved multiset $S$. Five metrics summarize the comparison. The intervention rate is defined across episodes, while the remaining four are computed within a single episode from its pair $(C,S)$ and then aggregated over episodes as described in Table~\ref{tab:action-simulator-audit}.
\begin{enumerate}
\item \textbf{Intervention rate:} the fraction of an agent's episodes in which the simulator changes the candidate list ($C\neq S$).
\item \textbf{Expansion ratio:} the length of the approved list relative to the candidate list, $|S|/|C|$.
\item \textbf{Candidate retention:} the share of the planner's original actions that survives approval,
\begin{equation}
\mathrm{Retention}(C,S)=\frac{\sum_a \min\{m_C(a),m_S(a)\}}{|C|},
\end{equation}
where $m_C(a)$ and $m_S(a)$ count the occurrences of action string $a$.
\item \textbf{Type preservation:} $|V(C)\cap V(S)|/|V(C)|$, where $V(\cdot)$ maps an action list to its set of substantive action verbs in production, trade, education, and work.
\item \textbf{Execution yield:} the fraction of simulator-approved actions completed before the first runtime correction or the next planning boundary. A runtime correction is either a clearing event, in which the runtime discards the remaining approved actions, or a replanning event, in which a new plan is issued. The associated clearing and replanning rates measure downstream correction pressure, not simulator rejection.
\end{enumerate}
The main audit covers the full architecture, which runs the complete planning--simulation--repair pipeline. Appendix~\ref{app:evaluation-details} reports planner-stratified summaries for the simplified variants as a robustness check.

Across E1--E4, the full architecture yields 10{,}149 paired planner--simulator episodes from 80 agents. The simulator changes most candidate lists, with agent-level intervention rates from 90.6\% in E3 to 100.0\% in E4 (Table~\ref{tab:action-simulator-audit}). These interventions are not wholesale replacements. In E1, type preservation reaches 95.2\% and candidate retention reaches 75.8\%, indicating that the simulator mainly expands and repairs plans while preserving the substantive production, trade, education, and work intent. E2 and E3 show the same pattern at lower preservation levels. E4 is the most constrained direct-production setting where the median expansion ratio is 5.56, candidate retention falls to 40.3\%, and runtime correction pressure remains high. Execution yield follows the same gradient, highest in E3 (73.7\%) and lowest in E4 (36.6\%), where clearing and replanning rates above 90\% show that the tiered re-planning path of Section~\ref{subsec:replanning} is exercised heavily under tight production constraints. Because each plan is validated against the state observed at planning time while the shared world keeps moving, residual infeasibility at execution is unavoidable. Therefore, the intended division of labor between pre-execution checking and runtime recovery is necessary.

\begin{table}[t]
\caption{Action Simulator trace audit for the full architecture in E1--E4. Rates are agent-level means with bootstrap 95\% confidence intervals; expansion is the episode-level median with interquartile range. Runtime correction reports action-list clearing and replanning rates after simulation.}
\label{tab:action-simulator-audit}
\centering
\scriptsize
\setlength{\tabcolsep}{3pt}
\resizebox{\textwidth}{!}{%
\begin{tabular}{@{}lrrrrrrr@{}}
\toprule
Experiment & Episodes & Changed (\%) & Expansion & Retained (\%) & Type kept (\%) & Yield (\%) & Clear/Replan (\%) \\
\midrule
E1 & 1{,}953 & 97.1 [96.4, 97.9] & 3.13 [2.11, 4.33] & 75.8 [74.5, 77.2] & 95.2 [94.5, 95.8] & 47.0 [45.2, 48.8] & 64.7 / 67.5 \\
E2 & 2{,}417 & 98.1 [97.4, 98.6] & 2.46 [1.63, 3.83] & 70.7 [69.2, 72.2] & 81.2 [78.8, 83.6] & 65.8 [64.5, 67.0] & 34.6 / 38.3 \\
E3 & 4{,}742 & 90.6 [89.8, 91.5] & 1.59 [1.00, 2.22] & 77.7 [76.2, 79.1] & 79.8 [77.9, 81.6] & 73.7 [72.6, 74.8] & 18.0 / 20.4 \\
E4 & 1{,}037 & 100.0 [100.0, 100.0] & 5.56 [4.17, 8.33] & 40.3 [38.5, 42.1] & 79.8 [78.3, 81.3] & 36.6 [35.3, 38.0] & 94.2 / 93.6 \\
\bottomrule
\end{tabular}
}
\end{table}

\begin{figure}[t]
\centering
\includegraphics[width=\textwidth]{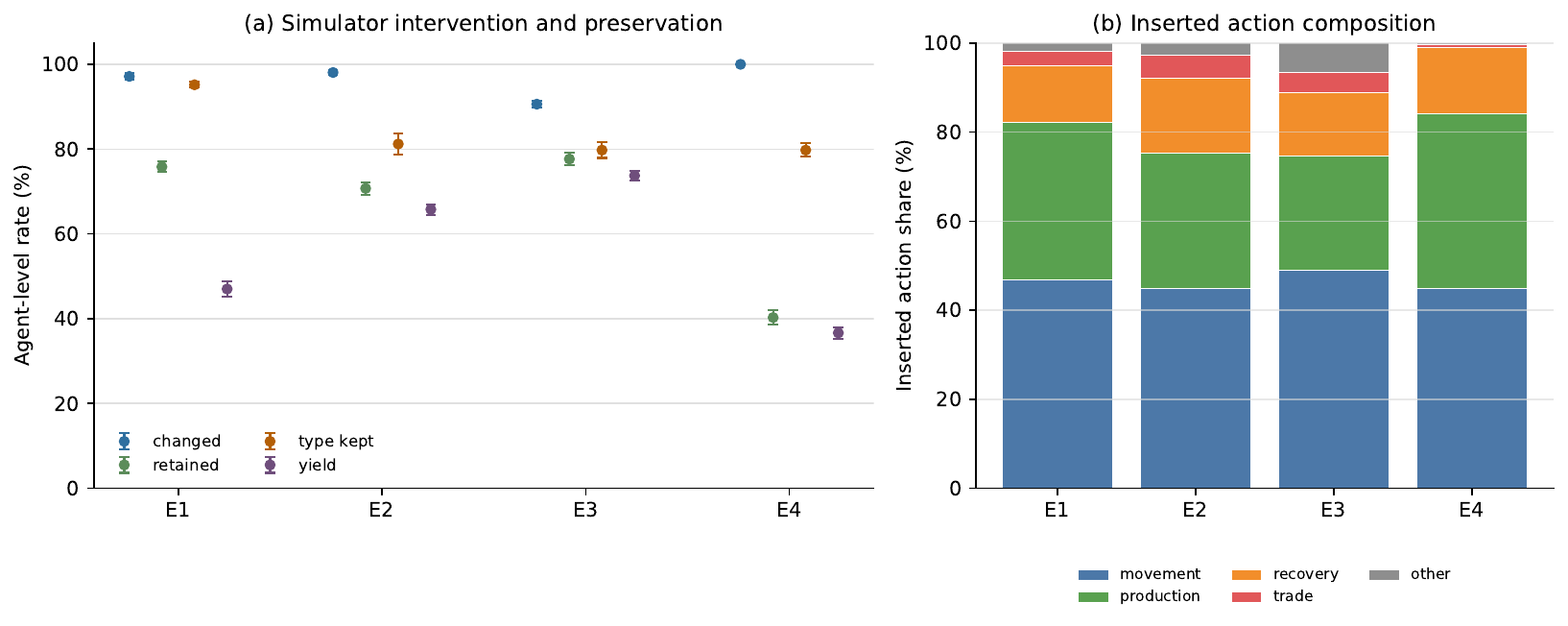}
\caption{Action Simulator trace audit for the full architecture. Panel A reports agent-level intervention, retention, type-preservation, and execution-yield rates with 95\% confidence intervals. Panel B decomposes inserted actions by category. The x-axis follows the controlled experiment codes in Table~\ref{tab:controlled-experiment-codes}.}
\label{fig:action-simulator-audit}
\end{figure}

The inserted-action composition further supports the intended architectural interpretation (Figure~\ref{fig:action-simulator-audit}). Movement, production, and recovery actions account for most inserted actions in all four tasks. These action classes correspond to the main sources of executable feasibility in the platform, since agents must reach the relevant location, obtain or produce prerequisites, and keep physiological state above execution thresholds.

We also label the constraint classes implicated by each simulator change, using the conservative proxy rules described in Appendix~\ref{app:evaluation-details}. In the full architecture, access or location, physiological, inventory or material, and budget or market proxies appear in 85.8\%, 79.0\%, 63.8\%, and 43.4\% of paired episodes, a residual task-logic label for deleted malformed actions appears in 2.7\%, and 83.6\% of episodes carry more than one label. Feasibility failures thus typically involve several constraint families at once, matching the coupled design of the environment. Despite the complexity of constraints, the simulator can still address most cases within a single intervention.

\subsection{Portrait and Steering Analyses}\label{subsec:portrait-protocol}

An agent's portrait is treated as persistent state that evolves through memory consolidation and carries human steering. The portrait analyses test whether this design holds in deployment, examining whether successive portrait updates form coherent identity trajectories and whether human steering leaves a measurable trace in them. The analyses use the raw public-deployment character-arc and human-interaction logs, covering 569{,}128 portrait-arc entries across 26{,}434 characters. Unless otherwise stated, statistics use agents with at least eight arcs ($N=13{,}173$).

\subsubsection{Portrait Temporal Coherence}\label{subsec:portrait-experiment}

Each portrait snapshot concatenates the six portrait fields and is represented within agent by character-level TF--IDF with bigrams and trigrams and a vocabulary cap of 5{,}000 features. The observational unit is the transition between two consecutive snapshots. Local temporal coherence is measured by an ordered within-agent contrast that compares each adjacent portrait pair with the next-nearest pair from the same starting portrait, and long-horizon change is measured by the early--late displacement, one minus the cosine similarity between the centroids of an agent's first three and last three portraits. The formal definitions are given in Appendix~\ref{app:evaluation-details}.

\begin{figure}[t]
\centering
\includegraphics[width=\linewidth]{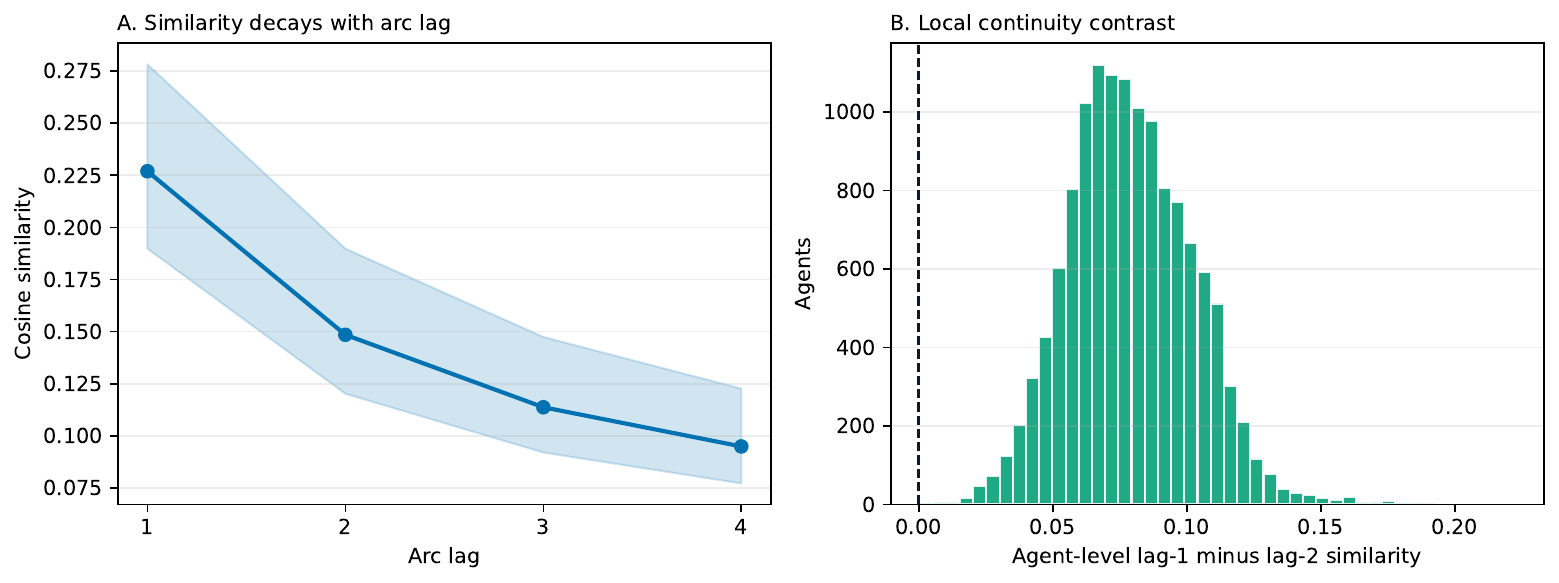}
\caption{Portrait temporal coherence in the public deployment ($N=13{,}173$ agents). Panel A shows the median agent-level similarity at arc lags one to four; shaded bands show the interquartile range. Panel B shows the distribution of the agent-level difference between lag-1 and lag-2 similarity, an agent-level view of the transition-level continuity contrast reported in the text.}
\label{fig:portrait-coherence}
\end{figure}

Across the 484{,}823 adjacent transitions, similarity decays consistently with arc lag. Mean within-agent cosine similarity is 0.238 for adjacent portraits, 0.159 at lag two, 0.123 at lag three, and 0.103 at lag four (Table~\ref{tab:portrait-coherence}; Figure~\ref{fig:portrait-coherence}). Absolute similarity levels are low by construction, since character-level TF--IDF vectors are sparse and high-dimensional. The evidence therefore rests on ordered within-agent comparisons and the monotonic decay indicates that successive updates retain local textual continuity.

\begin{table}[t]
\caption{Portrait temporal-coherence diagnostics ($N=13{,}173$ agents with at least eight arcs).}
\label{tab:portrait-coherence}
\centering
\small
\begin{tabular}{@{}lcccccc@{}}
\toprule
& \multicolumn{4}{c}{Similarity by arc lag} & Continuity & Early--late \\
\cmidrule(lr){2-5}
& 1 & 2 & 3 & 4 & contrast & displacement \\
\midrule
Mean & 0.238 & 0.159 & 0.123 & 0.103 & 0.092 & 0.891 \\
Median & 0.227 & 0.148 & 0.114 & 0.095 & --- & 0.897 \\
\bottomrule
\end{tabular}
\parbox{0.95\linewidth}{\footnotesize Lag similarities and early--late displacement are descriptive agent-level summaries; the continuity contrast is a transition-level estimate with two-way cluster-robust 95\% CI [0.088, 0.095].}
\end{table}

The ordered contrast confirms this pattern. Across 473{,}018 contrasts, adjacent portraits are more similar than next-nearest ones by 0.092 on average and 82.7\% of contrasts are positive. Early--late displacement remains large (mean 0.891), so portrait updates are locally coherent while remaining textually non-stationary over longer horizons, which is the signature of an identity that evolves without being rewritten.

\subsubsection{Human Steering and Portrait Updates}\label{subsec:steering-experiment}

\begin{table}[t]
	\caption{Adjusted within-agent associations between local human steering and adjacent portrait cosine distance. Positive estimates indicate larger short-horizon portrait updates relative to intervals without steering.}
	\label{tab:local-steering}
	\centering
	\small
	\begin{tabular}{@{}lrrr@{}}
		\toprule
		Local steering exposure & Estimate & 95\% CI & Holm-adjusted $p$ \\
		\midrule
		Any message & 0.0030 & [0.0011, 0.0048] & 0.0023 \\
		1 message & 0.0026 & [0.0003, 0.0049] & 0.058 \\
		2--3 messages & 0.0011 & [$-0.0010$, 0.0032] & 0.280 \\
		4--7 messages & 0.0040 & [0.0020, 0.0059] & $<0.001$ \\
		$\geq 8$ messages & 0.0084 & [0.0043, 0.0125] & $<0.001$ \\
		\bottomrule
	\end{tabular}
\end{table}

\begin{table}[t]
	\caption{Channel-specific within-agent associations with adjacent portrait cosine distance. Coefficients are estimated per unit increase in $\log(1+\text{count})$.}
	\label{tab:steering-channels}
	\centering
	\small
	\begin{tabular}{@{}lrrr@{}}
		\toprule
		Steering channel & Estimate & 95\% CI & Holm-adjusted $p$ \\
		\midrule
		Daily prompts & 0.0022 & [0.0006, 0.0037] & 0.021 \\
		Reactive commands & 0.0012 & [$-0.0001$, 0.0026] & 0.075 \\
		Free-form chat & 0.0026 & [0.0002, 0.0051] & 0.073 \\
		\bottomrule
	\end{tabular}
\end{table}

\begin{figure}[t]
	\centering
	\includegraphics[width=0.92\textwidth]{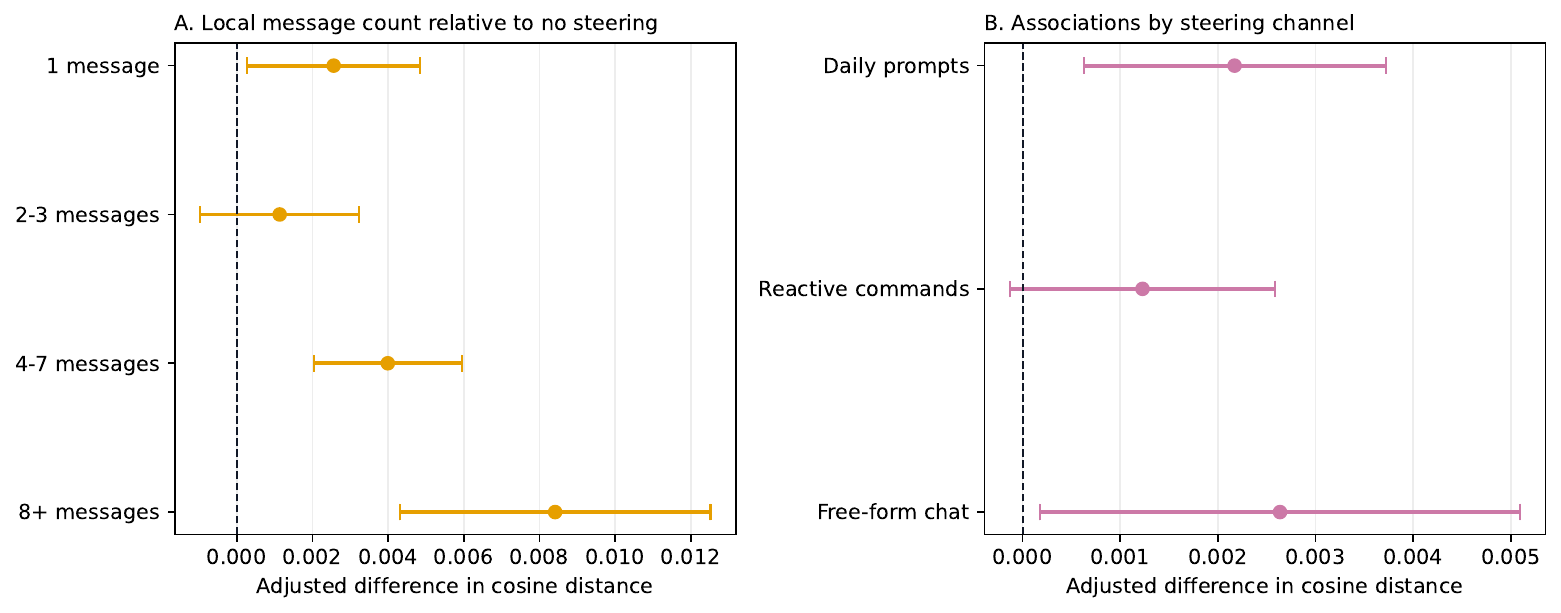}
	\caption{Adjusted within-agent associations between local human steering and adjacent portrait cosine distance. Panel A compares message-count categories with intervals without steering. Panel B shows channel-specific coefficients per unit increase in $\log(1+\text{count})$. Error bars show 95\% confidence intervals from two-way cluster-robust standard errors by agent and game day. Positive estimates indicate larger short-horizon portrait updates.}
	\label{fig:local-steering}
\end{figure}

This experiment quantifies the association between human steering and short-horizon portrait revision. Portrait revision over an interval is measured by the adjacent portrait distance $d_{a,t}$, which is defined by one minus the cosine similarity between agent $a$'s consecutive portrait vectors. Steering exposure is measured as the number of daily prompts, reactive commands, and free-form chat messages received in that interval. The association is estimated with the panel model
\begin{equation}
d_{a,t}=
\alpha_a+
\lambda_{\tau(t)}+
\boldsymbol{\beta}^{\top}\mathbf{z}_{a,t}+
\boldsymbol{\gamma}^{\top}\mathbf{x}_{a,t}+
\varepsilon_{a,t},
\label{eq:portrait-steering-panel}
\end{equation}
where $\alpha_a$ and $\lambda_{\tau(t)}$ are agent and game-day fixed effects, $\mathbf{z}_{a,t}$ encodes the steering exposure, and $\mathbf{x}_{a,t}$ collects interval-length and trajectory-position controls, with standard errors two-way cluster-robust by agent and game day. The design compares steered and unsteered intervals within the same agent. The exposure encodings and remaining specification details are given in Appendix~\ref{app:evaluation-details}. 

Human steering is associated with measurable short-horizon differences in portrait updates. Intervals containing at least one steering message have an adjusted portrait distance that is 0.0030 larger than intervals without steering. Across exposure levels, the estimates at one and at two to three messages are small and not individually supported, while intervals with four to seven messages show an adjusted difference of 0.0040 and intervals with at least eight messages show 0.0084, both supported after Holm adjustment (Table~\ref{tab:local-steering}; Figure~\ref{fig:local-steering}). The association therefore concentrates at higher local interaction densities, where steering is accompanied by progressively larger short-horizon portrait revision.

The channel-specific short-horizon model provides a more granular view (Table~\ref{tab:steering-channels}). After adjustment, daily prompts are associated with larger local portrait updates (0.0022). The corresponding command and chat estimates are positive but do not remain supported after correction. The channel pattern is consistent with the architecture's routing of steering, in which strategic input reshapes durable goal state while reactive commands and chat act through lighter, more local pathways.

\section{Conclusion}\label{sec:conclusion}

In this work, we presented AIvilization v0, a large-scale, persistent artificial society built around a central challenge of open-ended multi-agent simulation, reconciling teleological stability with reactive correctness under physiological constraints and market volatility. Moving beyond isolated chat-based interaction to a physically grounded, resource-constrained platform, AIvilization couples agents' survival, production, trade, education, and social life into a single closed loop.

First, we proposed a unified agent architecture that synergizes planning, memory, and execution. The Branch-Thinking Planner mitigates the fragility of long sequential plans by decomposing objectives into parallel branches, enabling agents to balance competing needs such as production, trading, and recovery, and within the planning structure the Action Simulator validates candidate actions against physical and economic constraints before execution and repairs the infeasible steps. The Adaptive Agent Profile separates immediate working context from slow semantic integration through dual-process memory, so agents react to real-time stimuli while accumulating social experience that gradually reshapes their decision priors and identity.

The controlled experiments validate these mechanisms. Removing branch decomposition or objective decomposition remains adequate for simple, well-specified tasks, in line with the lightweight planning route used for reactive human steering. In complex multi-objective settings, the full architecture is more robust and better balanced, most visibly in delayed-return investment, where it raises education outcomes over the ablated variants while managing production and physiological upkeep, and in open-ended exploration, where removing objective decomposition collapses action coverage. The trace audit complements these outcome-level results at the mechanism level. Across 10{,}149 planner--simulator episodes, the simulator intervenes in nearly every plan yet largely preserves the planner's intent, and most repairs address several constraint families at once. Besides, portrait updates formed locally coherent identity trajectories across 484{,}823 transitions, and denser human steering was associated with progressively larger short-horizon portrait revision, consistent with memory-mediated steering that adjusts identity without rewriting it.

Second, we engineered an environment that couples physiological survival constraints with an automated-market-maker economy and a gated education--occupation hierarchy. In place of predefined equilibrium assumptions, AIvilization enforces hard, non-substitutable production constraints and vertically structured supply chains, and this closed-loop design lets macroeconomic phenomena such as inflation and resource bottlenecks emerge from micro-level agent interactions.

A large-scale public deployment demonstrated the platform's empirical validity. The economy remained stable without freezing and spontaneously reproduced canonical stylized facts of real markets, heavy-tailed returns, volatility clustering, and volume--volatility co-movement across all 23 commodities, while realized prices were discovered through trading, anchored to but not dictated by the designed price ladder. Human-capital gating produced emergent stratification, with education and wealth strongly associated among employed agents. 

Finally, AIvilization operationalizes a hybrid ecosystem of human-guided and self-directed autonomy through memory-mediated steering. The evidence remains scoped to its design, with within-batch controlled comparisons and observational deployment findings, and extending it with across-seed replication and causal intervention designs is a natural next step. Taken together, the architecture, the environment, and the deployed evidence establish AIvilization as a testbed for studying long-horizon autonomy, emergent socioeconomic structure, and the collective dynamics of hybrid human--AI societies.

\section*{Acknowledgements}
This paper is the culmination of a large-scale project developed by the Bauhinia AI team. We are immensely proud of this collaborative effort and wish to extend our deepest gratitude to all who contributed to its success. The specific contributions are as follows (contributors are listed alphabetically by last name within each section):

\begin{itemize}
\item \textbf{Agent Architecture:} The core intelligent agents at the heart of our research were designed and implemented by Tsz Wai Chan, Xingyan Chen, Wenkai Fan, Xiaolong Wang, and Shurui Zhang. Their innovative work formed the technical backbone of this study.
\item \textbf{Platform Design:} The robust and user-friendly platform was designed and engineered by Xingyan Chen, Peiyan Xu, Haowei Yang, and Runjin Zhang. Their expertise was essential in creating a seamless experience for our participants.
\item \textbf{Game Environment and Illustration:} We are grateful to Junquan Bi, Jia Liu, Xiaolong Wang, Junming Zeng, and Shurui Zhang for developing the immersive game environment. The captivating illustrations, which brought our world to life, were created by Di Duan and Zhixuan Ouyang.
\item \textbf{Manuscript Preparation:} This paper was a collective writing effort. We acknowledge the significant contributions to the writing and revision process from Junquan Bi, Tsz Wai Chan, Xingyan Chen, Wenkai Fan, Jia Liu, Xiaolong Wang, Haowei Yang, Shurui Zhang, and Zirui Zhou.
\item \textbf{Business and Compliance:} We thank Zigeng Chen, Jianhao Xie, Haowei Yang, Shihua Zeng, and Zhonghui Zhang for their crucial work in navigating the business and compliance landscape, which provided the foundational support for our project's public launch.
\end{itemize}

We are profoundly indebted to Professor Kani Chen for his invaluable guidance, insightful discussions, and unwavering support throughout the entire lifecycle of this project.

Finally, and most importantly, we extend our heartfelt thanks to the tens of thousands of players who participated in our public experiment. This project would not have been possible without your enthusiastic engagement, time, and invaluable feedback. You were not just participants, but co-creators of this research, and we are incredibly grateful for your contribution.

\newpage
\begin{appendices}
\section{Agent Profile Example}\label{app:agent-profile}

Table~\ref{tab:agent-profile-example} shows a representative agent profile from the deployed platform, comprising the dynamic state, the long-term memory with its portrait fields and social records, and the short-term memory of recent execution traces.

\begin{table}[htbp]
\centering
\caption{An example of an agent profile.}
\label{tab:agent-profile-example}
\small
\renewcommand{\arraystretch}{1.2}
\begin{tabular}{@{}p{0.17\textwidth}p{0.2\textwidth}p{0.55\textwidth}@{}}
\toprule
\textbf{Category} & \textbf{Attribute} & \textbf{Content} \\
\midrule
\multirow{9}{=}{\textbf{Dynamic State}}
& Energy & 450 / 500 \\
& Satiety & 290 / 500 \\
& Health & 500 / 500 \\
& Education Score & 31 \\
& Current Balance & 191,696,904 \\
& Residential Tier & 5 \\
& Job & Stock Clerk \\
& Inventory & Transistor $\times$ 12; Iron Ore $\times$ 100; Copper Ore $\times$ 100; \\
&  & Fish $\times$ 46; Book $\times$ 1 \\
\midrule
\multirow{6}{=}{\textbf{Long-Term Memory}}
& Belief & Efficiency is paramount, with growing recognition of cooperation and flexibility. \\
& Mood & Predominantly rational, with increased patience and composure under change. \\
& Values & Coexistence of efficiency and symbiotic cooperation; pursuit of win--win outcomes. \\
& Habits & Prioritizes purchasing fish from the market when hungry; consistently engages in self-study after completing work tasks. \\
& Personality & MBTI: INTJ; reduced desire for control; improved coordination skills; transition from lone-wolf behavior toward team trust. \\
& Social Interaction Record &
Relation with Friend 1: Best Friends, Work Partner. \\
&  & Impression: Sustained cooperation has reinforced mutual trust and tacit understanding; detailed planning improved collaboration efficiency. \\
& & Relation with Friend 2: ... \\
\midrule
\multirow{3}{=}{\textbf{Short-Term Memory}}
& Successful Actions &
[craft Transistor 12, buy Fish 6, self-study 2, ...] \\
& Failed Actions &
craft Copper Ingot 12 \\
& & Reason: crafting Copper Ingot needs both Copper Ore and Wood, only Copper Ore in inventory \\
\bottomrule
\end{tabular}
\end{table}

\section{Environment Catalogs}\label{app:environment-catalogs}

This appendix reports the environment catalogs for the platform configuration described in Section~\ref{sec:environment}. Action entries are unit parameters unless otherwise stated. 

\begin{table}[htbp]
	\centering
	\caption{Executable unit actions with per-unit durations, costs, effects, and preconditions. E and S denote energy and satiety costs.}
	\label{tab:action-space}
	\scriptsize
	\setlength{\tabcolsep}{4pt}
	\begin{tabular}{@{}l>{\raggedright\arraybackslash}p{0.13\textwidth}>{\raggedright\arraybackslash}p{0.22\textwidth}>{\raggedright\arraybackslash}p{0.23\textwidth}>{\raggedright\arraybackslash}p{0.19\textwidth}@{}}
		\toprule
		Action & Duration (game h) & Cost per unit & Effect per unit & Precondition \\
		\midrule
		Sleep & 1 & -- & energy at tier rate$^{a}$ & -- \\
		Eat food & 0.05 & 1 food item & satiety by item$^{b}$ & item in inventory \\
		See doctor & 0.2 & 300 coins & $+10$ health & -- \\
		Paid learning & 1 & 200 coins, 20 E, 5 S & $+1$ education & residential access \\
		Reading & 1 & 1 book, 20 E, 5 S & $+1$ education & residential access \\
		Self-study & 3 & 60 E, 15 S & $+1$ education & -- \\
		Work & 1 & 20 E, 5 S & wage income$^{c}$ & employed \\
		Produce & recipe-specific$^{d}$ & recipe inputs and costs$^{d}$ & 1 output unit & residential access, materials \\
		Buy / Sell & immediate & AMM pricing$^{e}$ & inventory and balance change & market access \\
		\bottomrule
	\end{tabular}
	
	\parbox{0.95\textwidth}{\footnotesize
		$^{a}$~Tier-specific hourly recovery rate, Table~\ref{tab:residential-physiology}.
		$^{b}$~Item-specific recovery amount, Table~\ref{tab:food-recovery}.
		$^{c}$~Wage set by Equations~\eqref{eq:wage-static}--\eqref{eq:wage-dynamic}; base values in Table~\ref{tab:occupation-catalog}.
		$^{d}$~Per-unit material, energy, satiety, and time costs, Table~\ref{tab:production-recipes}.
		$^{e}$~Equations~\eqref{eq:amm-buy}--\eqref{eq:amm-sell}; fixed-value reward items sell at their configured value.}
\end{table}

\begin{table}[htbp]
\centering
\caption{Residential tiers and physiological capacity.}
\label{tab:residential-physiology}
\small
\begin{tabular}{@{}cccc@{}}
\toprule
Tier & Maximum energy/satiety/health & Sleep recovery per hour & Upgrade input \\
\midrule
1 & 80 & 10 & -- \\
2 & 200 & 25 & Copper Ore x 1 \\
3 & 320 & 40 & Iron Ore x 1 \\
4 & 360 & 45 & Silicon Ore x 1 \\
5 & 400 & 50 & Copper Ingot x 2 \\
\bottomrule
\end{tabular}
\end{table}

\begin{table}[htbp]
	\centering
	\caption{Satiety recovery from food consumption.}
	\label{tab:food-recovery}
	\small
	\begin{tabular}{@{}lrlrlr@{}}
		\toprule
		Food & Satiety & Food & Satiety & Food & Satiety \\
		\midrule
		Apple & 2 & Wheat & 12 & Rice & 16 \\
		Chicken & 32 & Beef & 48 & Fish & 60 \\
		Flour & 32 & Bread & 68 & Apple Pie & 74 \\
		Chicken Salad & 84 & Beef Rice & 104 & Sushi & 116 \\
		\bottomrule
	\end{tabular}
\end{table}

\begin{table}[htbp]
\centering
\caption{Commodity catalog with supply-chain tier, minimum residential tier requirement, and platform role.}
\label{tab:goods-catalog}
\scriptsize
\setlength{\tabcolsep}{3pt}
\resizebox{\textwidth}{!}{%
\begin{tabular}{@{}llcl@{}}
\toprule
Supply-chain tier & Commodity & $R_{\min}$ & Platform role \\
\midrule
Primary raw materials & Apple & 1 & Food processing input / basic consumption good \\
Primary raw materials & Wheat & 1 & Food processing input / basic consumption good \\
Primary raw materials & Rice & 1 & Food processing input / basic consumption good \\
Primary raw materials & Wood & 1 & Industrial raw material \\
Primary raw materials & Book & 1 & Basic consumption good \\
Primary raw materials & Copper Ore & 1 & Industrial raw material \\
Primary raw materials & Iron Ore & 1 & Industrial raw material \\
Primary raw materials & Silicon Ore & 1 & Industrial raw material \\
\midrule
Secondary processed food & Beef & 2 & Food processing input / secondary consumption good \\
Secondary processed food & Chicken & 2 & Food processing input / secondary consumption good \\
Secondary processed food & Fish & 3 & Food processing input / secondary consumption good \\
Secondary processed food & Flour & 3 & Food processing input / secondary consumption good \\
Secondary processed food & Bread & 3 & Secondary consumption good \\
Secondary processed food & Sushi & 3 & Secondary consumption good \\
Secondary processed food & Apple Pie & 3 & Secondary consumption good \\
Secondary processed food & Chicken Salad & 3 & Secondary consumption good \\
Secondary processed food & Beef Rice & 3 & Secondary consumption good \\
\midrule
Secondary refining materials & Coal & 4 & Industrial intermediate \\
Secondary refining materials & Copper Ingot & 4 & Industrial intermediate \\
Secondary refining materials & Iron Ingot & 4 & Industrial intermediate \\
Secondary refining materials & Pure Silicon & 4 & Industrial intermediate \\
\midrule
Tertiary high-tech industrial & Transistor & 5 & High-tech intermediate \\
Tertiary high-tech industrial & Circuit Board & 5 & High-tech intermediate \\
Tertiary high-tech industrial & Chip & 5 & High-tech consumption good \\
\midrule
Special reward & Gold Apple & -- & Rare byproduct; non-consumable, sellable at fixed value 10{,}000 \\
\bottomrule
\end{tabular}
}
\end{table}

\begin{table}[htbp]
\centering
\caption{Production recipes per unit. Energy and satiety are physiological costs; time is measured in platform game hours.}
\label{tab:production-recipes}
\scriptsize
\setlength{\tabcolsep}{3pt}
\resizebox{\textwidth}{!}{%
\begin{tabular}{@{}p{0.15\linewidth}p{0.34\linewidth}rrrr@{}}
\toprule
Output & Material inputs & Energy & Satiety & Time & Reward probability (\%) \\
\midrule
Apple & -- & 2 & 0 & 0.1 & 0 \\
Wheat & -- & 12 & 3 & 0.6 & 0 \\
Rice & -- & 16 & 4 & 0.8 & 0 \\
Wood & -- & 8 & 2 & 0.4 & 0 \\
Book & Wood x 1 & 32 & 8 & 1.6 & 0 \\
Copper Ore & -- & 12 & 3 & 0.6 & 0 \\
Iron Ore & -- & 16 & 4 & 0.8 & 0 \\
Silicon Ore & -- & 20 & 5 & 1 & 0 \\
\midrule
Beef & Wheat x 2 & 24 & 6 & 1.2 & 0 \\
Chicken & Wheat x 1 & 20 & 5 & 1 & 0 \\
Fish & -- & 60 & 15 & 3 & 0 \\
Flour & Wheat x 1 & 20 & 5 & 1 & 0 \\
Bread & Flour x 1 & 36 & 9 & 1.8 & 0 \\
Sushi & Rice x 1, Fish x 1 & 40 & 10 & 2 & 0 \\
Apple Pie & Apple x 1, Flour x 1 & 40 & 10 & 2 & 0 \\
Chicken Salad & Chicken x 1, Flour x 1 & 20 & 5 & 1 & 0.5 \\
Beef Rice & Rice x 1, Beef x 1 & 40 & 10 & 2 & 0.8 \\
\midrule
Coal & Wood x 1 & 40 & 10 & 2 & 0 \\
Copper Ingot & Wood x 1, Copper Ore x 1 & 48 & 12 & 2.4 & 0 \\
Iron Ingot & Iron Ore x 1, Coal x 1 & 52 & 13 & 2.6 & 0 \\
Pure Silicon & Silicon Ore x 1, Coal x 1 & 56 & 14 & 2.8 & 0 \\
\midrule
Transistor & Copper Ingot x 1, Iron Ingot x 1 & 60 & 15 & 3 & 1 \\
Circuit Board & Copper Ingot x 1, Pure Silicon x 1 & 80 & 20 & 4 & 2 \\
Chip & Transistor x 1, Circuit Board x 1 & 100 & 25 & 5 & 5 \\
\bottomrule
\end{tabular}
}
\end{table}

\begin{table}[htbp]
\centering
\caption{Job tier configuration under the default platform settings.}
\label{tab:job-tiers}
\scriptsize
\begin{tabular}{@{}clcclc@{}}
\toprule
Tier & Tier name & Min. $R$ & Min. $H$ & Prerequisite commodity & Wage type \\
\midrule
1 & Entry & 1 & 0 & -- & static \\
2 & Skilled & 2 & 20 & Beef & static \\
3 & Backbone & 3 & 70 & Sushi & static \\
4 & Expert & 4 & 110 & Pure Silicon & dynamic \\
5 & Management & 5 & 180 & Transistor & dynamic \\
6 & Leadership & 5 & 320 & Circuit Board & dynamic \\
\bottomrule
\end{tabular}
\end{table}

\begin{table}[htbp]
\centering
\caption{Occupation catalog and entry parameters under the default platform settings. The eligibility share $\pi_j$ sets the education requirement at the $(1-\pi_j)$-quantile of the current education distribution, floored at $H^{(j)}_{\mathrm{floor}}$ (Equation~\eqref{eq:dynamic-threshold}); $w^{(j)}_0$ is the base wage before price-index scaling.}
\label{tab:occupation-catalog}
\scriptsize
\begin{tabular}{@{}lccccc@{}}
\toprule
Occupation & Tier & $R^{(j)}_{\min}$ & $H_{\mathrm{floor}}^{(j)}$ & $\pi_j$ & $w^{(j)}_0$ \\
\midrule
Cleaner & 1 & 1 & 0 & 1.00 & 250 \\
Waiter & 1 & 1 & 13 & 0.90 & 253 \\
\midrule
Stock Clerk & 2 & 2 & 0 & 0.832 & 260 \\
Security Guard & 2 & 2 & 42 & 0.728 & 270 \\
Receptionist & 2 & 2 & 62 & 0.624 & 275 \\
\midrule
Cashier & 3 & 3 & 78 & 0.560 & 301 \\
Maintenance Worker & 3 & 3 & 104 & 0.476 & 309 \\
\midrule
Chef & 4 & 4 & 113 & 0.448 & 356 \\
Nurse & 4 & 4 & 141 & 0.384 & 366 \\
Teacher & 4 & 4 & 176 & 0.320 & 380 \\
\midrule
Doctor & 5 & 5 & 207 & 0.280 & 429 \\
Office Clerk & 5 & 5 & 237 & 0.245 & 444 \\
Supermarket Manager & 5 & 5 & 273 & 0.210 & 463 \\
Restaurant Manager & 5 & 5 & 319 & 0.175 & 489 \\
\midrule
Principal & 6 & 5 & 357 & 0.150 & 734 \\
Hospital Director & 6 & 5 & 421 & 0.120 & 961 \\
CEO & 6 & 5 & 604 & 0.065 & 1411 \\
\bottomrule
\end{tabular}
\end{table}

\FloatBarrier
\section{Additional Evaluation Details}\label{app:evaluation-details}

This appendix records the conventions, robustness summaries, and full model specifications supporting the protocols of Section~\ref{sec:evaluation-protocol}. 
\subsection{Public-Deployment Economy Analyses}\label{app:economy-details}

This subsection defines the market indicators used in Section~\ref{subsec:market-experiment}. Throughout, a commodity's five-minute close prices are $P_t$, $t=1,\ldots,T$, with log prices $\ell_t=\log P_t$, log returns $r_t=\ell_t-\ell_{t-1}$, and traded volumes $V_t$.

\paragraph{Log-price range.} The range measures the total extent of price variation on the logarithmic scale,
\begin{equation}
\mathrm{Range}=\max_{1\le t\le T}\ell_t-\min_{1\le t\le T}\ell_t .
\label{eq:app-range}
\end{equation}
A path with unbounded growth or a collapse toward zero produces a large range, so a small value over a long horizon indicates the absence of explosive or degenerate price dynamics.

\paragraph{Maximum drawdown.} The drawdown at time $t$ is the fractional loss relative to the running peak, and its maximum
\begin{equation}
\mathrm{MDD}=\max_{1\le t\le T}\left(1-\frac{P_t}{\max_{1\le s\le t}P_s}\right)
\label{eq:app-mdd}
\end{equation}
records the most severe peak-to-trough decline in the window. Crash-like episodes register here even when the series later recovers, which an endpoint comparison would miss.

\paragraph{Excess kurtosis and skewness.} Excess kurtosis is the standardized fourth central moment of $r_t$ minus three, and skewness is the standardized third central moment. A Gaussian distribution has excess kurtosis zero, so positive values indicate that extreme returns occur more often than a normal benchmark predicts, and nonzero skewness indicates asymmetry between large upward and large downward moves.

\paragraph{Autocorrelation of absolute returns and volatility memory.} Absolute returns proxy the magnitude of price movement, and their autocorrelation
\begin{equation}
\rho(k)=\frac{\sum_{t=k+1}^{T}\big(|r_t|-\overline{|r|}\big)\big(|r_{t-k}|-\overline{|r|}\big)}{\sum_{t=1}^{T}\big(|r_t|-\overline{|r|}\big)^{2}},
\qquad k\ge 1,
\label{eq:app-acf}
\end{equation}
where $\overline{|r|}$ is the sample mean of $|r_t|$, is positive when volatile periods cluster in time. The lag-1 value captures immediate persistence, and the volatility-memory score $A=\sum_{k=1}^{10}\rho(k)$ accumulates persistence over the first ten lags into a single summary.

\paragraph{Ljung--Box test.} The test aggregates the first ten autocorrelations of $|r_t|$ into one statistic,
\begin{equation}
Q=T(T+2)\sum_{k=1}^{10}\frac{\hat{\rho}(k)^{2}}{T-k},
\label{eq:app-lb}
\end{equation}
referred to a $\chi^{2}_{10}$ null distribution \cite{ljung1978measure}. Rejection indicates that the observed volatility clustering is jointly significant across lags, not an artifact of any single lag.

\paragraph{Volume--volatility association.} The correlation $\mathrm{corr}\big(|r_t|,\log(1+V_t)\big)$ relates the magnitude of price changes to log-scaled traded volume. Positive values indicate that trading activity intensifies when prices move strongly, a co-movement regularity documented for real markets.

\paragraph{Realized growth.} The ratio $G=P_T/P_0$ compares the final and initial prices. It summarizes long-run drift over the window and supports comparison of price trajectories across supply-chain positions.

\begin{table}[htbp]
\centering
\caption{Supply-chain group diagnostics in the public-deployment commodity markets.}
\label{tab:app-environment-group-diagnostics}
\small
\begin{tabular}{@{}lrrrr@{}}
\toprule
Group & Items & Growth & Excess kurt. & ACF sum \\
\midrule
Raw food & 4 & 1.54 & 141.0 & 1.52 \\
Raw material & 5 & 0.98 & 1708.0 & 1.74 \\
Processed goods & 9 & 1.79 & 137.0 & 1.27 \\
Refined material & 3 & 1.76 & 42.5 & 0.50 \\
Industrial & 2 & 10.03 & 30.8 & 1.26 \\
\bottomrule
\end{tabular}
\parbox{0.95\linewidth}{\footnotesize Values are medians over commodities within each group. Growth is $P_T/P_0$; ACF sum is $\sum_{\ell=1}^{10}\mathrm{corr}(|r_t|,|r_{t-\ell}|)$.}
\end{table}

\begin{table}[htbp]
\centering
\caption{Observed education--occupation--wealth sorting among employed characters.}
\label{tab:app-occupation-sorting}
\scriptsize
\begin{tabular}{@{}lrrrr@{}}
\toprule
Occupation & Agents & Median wealth & Mean wealth & Median education \\
\midrule
Cleaner & 3,519 & 0.026 & 0.430 & 20.0 \\
Waiter & 2,122 & 0.064 & 0.194 & 39.0 \\
Stock Clerk & 628 & 0.120 & 2.036 & 32.0 \\
Security Guard & 837 & 0.163 & 0.778 & 46.0 \\
Receptionist & 952 & 0.251 & 1.414 & 70.0 \\
Cashier & 768 & 0.466 & 2.538 & 87.0 \\
Maintenance Worker & 449 & 0.594 & 4.077 & 130.0 \\
Chef & 447 & 1.082 & 5.546 & 171.0 \\
Nurse & 239 & 1.159 & 12.198 & 215.0 \\
Teacher & 527 & 1.365 & 9.093 & 316.0 \\
Doctor & 141 & 2.319 & 16.889 & 345.0 \\
Office Clerk & 243 & 2.582 & 10.246 & 445.0 \\
Supermarket Manager & 171 & 3.117 & 9.008 & 453.0 \\
Restaurant Manager & 179 & 4.520 & 13.165 & 563.0 \\
Principal & 370 & 6.662 & 19.199 & 1025.5 \\
Hospital Director & 88 & 6.512 & 53.993 & 1005.0 \\
CEO & 176 & 7.972 & 33.418 & 1293.0 \\
\bottomrule
\end{tabular}
\parbox{0.95\textwidth}{\footnotesize Wealth values are reported in millions of coins. The table is descriptive and excludes unemployed characters.}
\end{table}

\subsection{Controlled Ablation}

The contrast tables in this subsection report four statistics for each pairwise comparison, as strengthened statistical support for the results of Section~\ref{subsec:controlled-protocol}. The mean difference is the full-architecture mean minus the comparison-variant mean. Its percentile-bootstrap 95\% confidence interval is obtained by resampling agents with replacement within each batch 10{,}000 times and taking the 2.5th and 97.5th percentiles of the resampled differences. Significance is assessed with a two-sided Mann--Whitney $U$ test on the per-agent values, and Holm adjustment is applied within each task-level family of contrasts. Effect sizes are reported as Cliff's $\delta$, the probability that a randomly drawn agent from one variant exceeds a randomly drawn agent from the other minus the reverse probability. Throughout, time-aligned quantities use each agent's active execution time, the cumulative interval between consecutive logged events with idle gaps longer than four hours excluded.

\begin{table}[htbp]
\centering
\caption{Pairwise contrasts for E1 ($n=80$ agents per planner variant).}
\label{tab:app-complex-contrasts}
\scriptsize
\setlength{\tabcolsep}{4pt}
\resizebox{\textwidth}{!}{%
\begin{tabular}{@{}llrrr@{}}
\toprule
Comparison & Outcome & Mean difference [bootstrap 95\% CI] & Cliff's $\delta$ & Holm-adjusted $p$ \\
\midrule
Full architecture vs Without-Branch & Net worth & 34{,}807 [18{,}001, 52{,}668] & 0.17 & 0.356 \\
Full architecture vs Without-Branch & Education & 14.51 [11.75, 17.28] & 0.77 & $<10^{-8}$ \\
Full architecture vs Without-Branch & High-value item count & 0.45 [$-1.61$, 2.60] & $-0.04$ & 1.000 \\
Full architecture vs Without-Branch & Successful action events & 11.89 [$-1.39$, 25.15] & 0.26 & 0.034 \\
\midrule
Full architecture vs Without-OD & Net worth & 14{,}384 [$-3{,}495$, 33{,}085] & $-0.05$ & 1.000 \\
Full architecture vs Without-OD & Education & 16.80 [14.38, 19.29] & 0.89 & $<10^{-8}$ \\
Full architecture vs Without-OD & High-value item count & 0.28 [$-1.63$, 2.25] & $-0.10$ & 1.000 \\
Full architecture vs Without-OD & Successful action events & $-11.50$ [$-24.99$, 1.99] & $-0.07$ & 1.000 \\
\bottomrule
\end{tabular}
}
\end{table}

\begin{table}[htbp]
\centering
\caption{Pairwise contrasts for E2 ($n=80$ agents per planner variant).}
\label{tab:app-wealth-contrasts}
\scriptsize
\setlength{\tabcolsep}{4pt}
\resizebox{\textwidth}{!}{%
\begin{tabular}{@{}llrrr@{}}
\toprule
Comparison & Outcome & Mean difference [bootstrap 95\% CI] & Cliff's $\delta$ & Holm-adjusted $p$ \\
\midrule
Full architecture vs Without-Branch & Inventory value & 32{,}697 [24{,}803, 40{,}944] & 0.66 & $<10^{-8}$ \\
Full architecture vs Without-Branch & Net worth & 76{,}211 [31{,}683, 121{,}408] & 0.24 & 0.037 \\
Full architecture vs Without-Branch & Education & 21.71 [10.00, 33.76] & 0.31 & 0.003 \\
\midrule
Full architecture vs Without-OD & Inventory value & 24{,}486 [16{,}253, 32{,}881] & 0.50 & $<10^{-6}$ \\
Full architecture vs Without-OD & Net worth & $-19{,}460$ [$-67{,}086$, 28{,}705] & $-0.13$ & 0.305 \\
Full architecture vs Without-OD & Education & $-33.14$ [$-47.30$, $-18.99$] & $-0.38$ & $<0.001$ \\
\bottomrule
\end{tabular}
}
\end{table}

\begin{table}[htbp]
\centering
\caption{Pairwise contrasts for E3 ($n=80$ agents per planner variant).}
\label{tab:app-diversity-contrasts}
\scriptsize
\setlength{\tabcolsep}{4pt}
\resizebox{\textwidth}{!}{%
\begin{tabular}{@{}llrrr@{}}
\toprule
Comparison & Outcome & Mean difference [bootstrap 95\% CI] & Cliff's $\delta$ & Holm-adjusted $p$ \\
\midrule
Full architecture vs Without-Branch & Total unique actions & 0.59 [$-1.54$, 2.48] & 0.06 & 0.992 \\
Full architecture vs Without-OD & Total unique actions & 23.95 [21.63, 26.24] & 0.98 & $<10^{-8}$ \\
\bottomrule
\end{tabular}
}
\end{table}

\begin{table}[htbp]
\centering
\caption{Pairwise contrasts for E4 ($n=80$ agents per planner variant).}
\label{tab:app-chip-contrasts}
\scriptsize
\setlength{\tabcolsep}{4pt}
\resizebox{\textwidth}{!}{%
\begin{tabular}{@{}llrrr@{}}
\toprule
Comparison & Outcome & Mean difference [bootstrap 95\% CI] & Cliff's $\delta$ & Holm-adjusted $p$ \\
\midrule
Full architecture vs Without-Branch & Manufactured chips & $-0.01$ [$-0.93$, 0.93] & $-0.03$ & 0.741 \\
Full architecture vs Without-OD & Manufactured chips & 0.54 [$-0.31$, 1.41] & 0.08 & 0.741 \\
\bottomrule
\end{tabular}
}
\end{table}

\subsection{Action Simulator Trace Audit}

Table~\ref{tab:app-action-simulator-robustness} reports the audit metrics stratified by planner variant, complementing the full-architecture audit of Section~\ref{subsec:audit-experiment}.

\begin{table}[htbp]
\centering
\caption{Planner-stratified audit metrics for E1--E4 (agent-level means).}
\label{tab:app-action-simulator-robustness}
\scriptsize
\setlength{\tabcolsep}{3pt}
\resizebox{\textwidth}{!}{%
\begin{tabular}{@{}llrrrrrr@{}}
\toprule
Experiment & Planner & Episodes & Changed (\%) & Expansion & Retained (\%) & Type kept (\%) & Yield (\%) \\
\midrule
E1 & Full architecture & 1{,}953 & 97.1 & 3.13 & 75.8 & 95.2 & 47.0 \\
E1 & Without-Branch & 1{,}604 & 99.9 & 3.33 & 64.4 & 87.3 & 44.0 \\
E1 & Without-OD & 1{,}965 & 97.8 & 2.83 & 78.9 & 94.9 & 59.4 \\
E2 & Full architecture & 2{,}417 & 98.1 & 2.46 & 70.7 & 81.2 & 65.8 \\
E2 & Without-Branch & 1{,}943 & 99.9 & 2.00 & 66.9 & 84.4 & 73.0 \\
E2 & Without-OD & 4{,}086 & 94.2 & 2.00 & 77.0 & 81.5 & 65.5 \\
E3 & Full architecture & 4{,}742 & 90.6 & 1.59 & 77.7 & 79.8 & 73.7 \\
E3 & Without-Branch & 3{,}587 & 99.3 & 1.63 & 87.1 & 96.0 & 81.6 \\
E3 & Without-OD & 6{,}337 & 66.3 & 1.16 & 93.9 & 96.1 & 77.7 \\
E4 & Full architecture & 1{,}037 & 100.0 & 5.56 & 40.3 & 79.8 & 36.6 \\
E4 & Without-Branch & 1{,}515 & 99.9 & 4.17 & 62.4 & 79.6 & 47.5 \\
E4 & Without-OD & 1{,}074 & 100.0 & 15.83 & 52.9 & 99.9 & 34.6 \\
\bottomrule
\end{tabular}
}
\end{table}

The constraint-class labels of Section~\ref{subsec:audit-experiment} are assigned by deterministic keyword rules over the actions the simulator inserts or deletes in each paired episode. An episode is labeled physiological when recovery actions (sleeping, eating, or hospital visits) are inserted; inventory or material when production actions are inserted; budget or market when trade actions are inserted or a trade quantity is changed; access or location when movement actions are inserted; and task logic when malformed candidate actions, those that are empty, exceed four tokens, or carry an unrecognized verb, are deleted. The labels are proxies derived from repair behavior, not manual causal annotations. Table~\ref{tab:app-proxy-taxonomy} reports the resulting label counts.

\begin{table}[htbp]
\centering
\caption{Constraint-class proxy labels in the full-architecture audit (10{,}149 paired episodes).}
\label{tab:app-proxy-taxonomy}
\scriptsize
\setlength{\tabcolsep}{4pt}
\begin{tabular}{@{}lccccc@{}}
\toprule
& Access/location & Physiological & Inventory/material & Budget/market & Task logic \\
\midrule
Episodes & 8{,}709 & 8{,}018 & 6{,}479 & 4{,}409 & 271  \\
Share (\%) & 85.8 & 79.0 & 63.8 & 43.4 & 2.7 \\
\bottomrule
\end{tabular}
\end{table}

\subsection{Portrait and Steering Models}

Each character arc concatenates the six portrait fields, belief, mood, values, habits, personality, and diary, and is transformed within agent by the TF--IDF representation of Section~\ref{subsec:portrait-experiment}. For an ordered sequence of portrait vectors $\mathbf{v}_{a,1},\ldots,\mathbf{v}_{a,L}$ for agent $a$, the lag-$k$ similarity is $\cos(\mathbf{v}_{a,t},\mathbf{v}_{a,t+k})$. The local temporal-coherence contrast is
\begin{equation}
g_{a,t}=
\cos(\mathbf{v}_{a,t},\mathbf{v}_{a,t+1})
-
\cos(\mathbf{v}_{a,t},\mathbf{v}_{a,t+2}).
\label{eq:portrait-local-coherence}
\end{equation}
Its mean is estimated with an intercept-only linear model and two-way cluster-robust standard errors by agent and game day. Early--late displacement is $1-\cos(\bar{\mathbf{v}}_{a,\mathrm{early}},\bar{\mathbf{v}}_{a,\mathrm{late}})$, where the centroids use the first and last three arcs.

In the panel model of Equation~\eqref{eq:portrait-steering-panel}, steering events are aligned to the real-time interval between two consecutive portrait snapshots; $d_{a,t}=1-\cos(\mathbf{v}_{a,t},\mathbf{v}_{a,t+1})$ is the adjacent portrait distance, $\mathbf{z}_{a,t}$ encodes local steering exposure, $\mathbf{x}_{a,t}$ contains the log-transformed game-time interval, real-time interval, and arc index, and $\tau(t)=\lfloor \mathrm{game\_time}/24 \rfloor$ indexes game days. We estimate separate specifications for any steering event, local message-count categories (0, 1, 2--3, 4--7, and $\geq 8$), and log-transformed counts for daily prompts, reactive commands, and free-form chat. Holm adjustment is applied within each reported coefficient family.

\end{appendices}

\newpage
\bibliographystyle{abbrvnat}
\bibliography{references}

\end{document}